\documentclass[a4paper, oneside, twocolumn, notitlepage, 10pt]{extarticle_ecoc}
\usepackage{ecoc}

\usepackage{amsmath,amssymb}
\usepackage{pgfplotstable}
\usepackage{subcaption}
\usepackage{siunitx}
\usepackage{url}
\definecolor{C0}{RGB}{031, 119, 180} 
\definecolor{C1}{RGB}{255, 127, 014} 
\definecolor{C2}{RGB}{044, 160, 044} 
\definecolor{C3}{RGB}{215, 039, 040} 
\definecolor{C4}{RGB}{148, 103, 189} 
\definecolor{C5}{RGB}{140, 086, 075} 
\definecolor{C6}{RGB}{227, 119, 194} 
\definecolor{C7}{RGB}{127, 127, 127} 
\definecolor{C8}{RGB}{188, 189, 034} 
\definecolor{C9}{RGB}{023, 190, 207} 

\definecolor{C10}{RGB}{174, 199, 232} 
\definecolor{C11}{RGB}{255, 187, 120} 
\definecolor{C12}{RGB}{152, 223, 138} 
\definecolor{C13}{RGB}{255, 152, 150} 
\definecolor{C14}{RGB}{197, 176, 213} 
\definecolor{C15}{RGB}{196, 156, 148} 
\definecolor{C16}{RGB}{247, 182, 210} 
\definecolor{C17}{RGB}{199, 199, 199} 
\definecolor{C18}{RGB}{219, 219, 141} 
\definecolor{C19}{RGB}{158, 218, 229} 
\usepackage[font=small]{caption}
 \usetikzlibrary{backgrounds}
\usepackage{import}
\usepackage{titlesec}
\usepackage{tikz}
\usepackage{import}

\usetikzlibrary{positioning}
\usetikzlibrary{fit}
\usetikzlibrary{calc}
\usetikzlibrary{shapes}
\usetikzlibrary{math}
\usetikzlibrary{fpu}
\usetikzlibrary{decorations.markings}
\usetikzlibrary{decorations.pathreplacing}
\usetikzlibrary{arrows.meta}
\usetikzlibrary{intersections}

\makeatletter

\tikzset{outer sep=0}
\tikzset{inner sep=0}

\def\pgfaddtoshape#1#2{
	\begingroup
	\def\pgf@sm@shape@name{#1}%
	\let\anchor\pgf@sh@anchor
	#2%
	\endgroup
}

\newcommand{\anchorlet}[2]{
	\global\expandafter
	\let\csname pgf@anchor@\pgf@sm@shape@name @#1\expandafter\endcsname
	\csname pgf@anchor@\pgf@sm@shape@name @#2\endcsname
}

\pgfmathsetmacro{\NODESIZE}{42}
\pgfmathsetmacro{\NODETHICKNESS}{1.0}
\pgfmathsetmacro{\ROUNDEDCORNERS}{0.5mm}
\tikzset{node distance=0.5*\NODESIZE pt}

\def\FNODESIZE{\NODESIZE pt}

\tikzstyle{textstyle} = [text height=1.5ex, text depth=.5ex]
\tikzset{every label/.style=textstyle}

\tikzstyle{linestyle} = [line width = \NODETHICKNESS, rounded corners = \ROUNDEDCORNERS]
\tikzstyle{arrowstyle} = [>=stealth, linestyle]
\tikzstyle{<--} = [<-, arrowstyle]
\tikzstyle{-->} = [->, arrowstyle]
\tikzstyle{->-} = [linestyle, decoration={markings,	mark=at position 0.5 with {\arrow[arrowstyle]{>}}}, postaction={decorate}] 
\tikzstyle{-<-} = [linestyle, decoration={markings,	mark=at position 0.5 with {\arrow[arrowstyle]{<}}}, postaction={decorate}] 

\tikzset{   
    -A-/.style args={#1}{%
        linestyle, 
        decoration={markings, mark=at position #1 with {\arrow[>=Triangle, scale=.025*\NODESIZE]{>}}},
        postaction={decorate}
    },
    -A-/.default = {0.75}
}

\tikzset{   
    -AA-/.style args={#1, #2, #3}{%
        linestyle,
        decoration={markings, mark=at position #1 with {
            \node[amp={color=#2, width=#3, height=#3}, rotate=\pgfdecoratedangle] at (0,0) (inline_amp) {};
        }},
        postaction={decorate}
    },
    -AA-/.default = {0.5, E, 0.15}
}

\tikzstyle{<-->} = [<->, arrowstyle]
\tikzstyle{---} = [arrowstyle]

\tikzset{
    -PC-/.style args={#1}{
        linestyle, 
        decoration={markings, mark=at position #1 with {
            \draw[---,FO]
                (.05*\NODESIZE*\pgflinewidth,.05*\NODESIZE*\pgflinewidth) circle[radius=.05*\NODESIZE*\pgflinewidth];
            \draw[---,FO]
                (-.05*\NODESIZE*\pgflinewidth,.05*\NODESIZE*\pgflinewidth) circle[radius=.05*\NODESIZE*\pgflinewidth];
            \draw[---,FO]
                (0,-.05*\NODESIZE*\pgflinewidth) circle[radius=.05*\NODESIZE*\pgflinewidth];}},
        postaction={decorate}
        },
        -PC-/.default = {0.5}
}

\definecolor{C0}{RGB}{031, 119, 180} 
\definecolor{C1}{RGB}{031, 119, 180} 
\definecolor{C2}{RGB}{255, 127, 014} 
\definecolor{C3}{RGB}{044, 160, 044} 
\definecolor{C4}{RGB}{215, 039, 040} 
\definecolor{C6}{RGB}{148, 103, 189} 
\definecolor{C100}{RGB}{140, 086, 075} 
\definecolor{C7}{RGB}{227, 119, 194} 
\definecolor{C8}{RGB}{127, 127, 127} 
\definecolor{C9}{RGB}{188, 189, 034} 
\definecolor{C5}{RGB}{023, 190, 207} 

\definecolor{C0l}{RGB}{174, 199, 232} 
\definecolor{C11}{RGB}{174, 199, 232} 
\definecolor{C12}{RGB}{255, 187, 120} 
\definecolor{C13}{RGB}{152, 223, 138} 
\definecolor{C14}{RGB}{255, 152, 150} 
\definecolor{C15}{RGB}{197, 176, 213} 
\definecolor{C16}{RGB}{196, 156, 148} 
\definecolor{C17}{RGB}{247, 182, 210} 
\definecolor{C18}{RGB}{199, 199, 199} 
\definecolor{C19}{RGB}{219, 219, 141} 
\definecolor{C20}{RGB}{158, 218, 229} 

\definecolor{Snow}{HTML}{FBFBFB} 			
\definecolor{TUeRed}{RGB}{200, 25, 25}		
\definecolor{TUeGreen}{RGB}{25, 200, 113}	
\definecolor{TUeBlue}{RGB}{25, 113, 200}	

\definecolor{O}{RGB}{031, 119, 180} 	
\definecolor{Ol}{RGB}{174, 199, 232} 	
\definecolor{E}{RGB}{255, 127, 014} 	
\definecolor{El}{RGB}{255, 187, 120} 	
\definecolor{D}{RGB}{148, 103, 189}     
\definecolor{Dl}{RGB}{197, 176, 213} 	
\definecolor{EO}{RGB}{215, 039, 040} 	
\definecolor{EOl}{RGB}{255, 152, 150}   

\tikzstyle{FW} = [fill=white]			
\tikzstyle{FB} = [fill=white]			
\tikzstyle{FO} = [fill=C0l, draw=C0]	
\tikzstyle{FE} = [fill=C1l, draw=C1]	
\tikzstyle{FD} = [fill=C2l, draw=C2]	

\def\direce{e}
\def\direcw{w}
\def\direcn{n}
\def\direcs{s}

\def\flipfalse{0}
\newcount\portcount
\pgfdeclareshape{basic}{
	\inheritsavedanchors[from=rectangle] 
	
	\inheritanchor[from=rectangle]{center}
	\inheritanchor[from=rectangle]{mid}		
	\inheritanchor[from=rectangle]{base}	
	\inheritanchor[from=rectangle]{north}
	\inheritanchor[from=rectangle]{south}		
	\inheritanchor[from=rectangle]{west}		
	\inheritanchor[from=rectangle]{mid west}				
	\inheritanchor[from=rectangle]{base west}		
	\inheritanchor[from=rectangle]{north west}		
	\inheritanchor[from=rectangle]{south west}		
	\inheritanchor[from=rectangle]{east}
	\inheritanchor[from=rectangle]{mid east}
	\inheritanchor[from=rectangle]{base east}	
	\inheritanchor[from=rectangle]{north east}
	\inheritanchor[from=rectangle]{south east}	
	
	\inheritanchorborder[from=rectangle]	

	\inheritbackgroundpath[from=rectangle]	
}

\pgfaddtoshape{basic}{
	\anchor{west south west}{
		\pgf@process{\northeast}
		\pgf@ya=.5\pgf@y
		\pgf@process{\southwest}
		\pgf@y=1.5\pgf@y
		\advance\pgf@y by \pgf@ya
		\pgf@y=.5\pgf@y
	}
	\anchor{west north west}{
		\pgf@process{\northeast}
		\pgf@ya=1.5\pgf@y
		\pgf@process{\southwest}
		\pgf@y=.5\pgf@y
		\advance\pgf@y by \pgf@ya
		\pgf@y=.5\pgf@y
	}
	\anchor{east north east}{
		\pgf@process{\southwest}
		\pgf@ya=.5\pgf@y
		\pgf@process{\northeast}
		\pgf@y=1.5\pgf@y
		\advance\pgf@y by \pgf@ya
		\pgf@y=.5\pgf@y
	}
	\anchor{east south east}{
		\pgf@process{\southwest}
		\pgf@ya=1.5\pgf@y
		\pgf@process{\northeast}
		\pgf@y=.5\pgf@y
		\advance\pgf@y by \pgf@ya
		\pgf@y=.5\pgf@y
	}
	\anchor{north north west}{
		\pgf@process{\southwest}
		\pgf@xa=1.5\pgf@x
		\pgf@process{\northeast}
		\pgf@x=.5\pgf@x
		\advance\pgf@x by \pgf@xa
		\pgf@x=.5\pgf@x
	}
	\anchor{north north east}{
		\pgf@process{\southwest}
		\pgf@xa=.5\pgf@x
		\pgf@process{\northeast}
		\pgf@x=1.5\pgf@x
		\advance\pgf@x by \pgf@xa
		\pgf@x=.5\pgf@x
	}
	\anchor{south south west}{
		\pgf@process{\northeast}
		\pgf@xa=.5\pgf@x
		\pgf@process{\southwest}
		\pgf@x=1.5\pgf@x
		\advance\pgf@x by \pgf@xa
		\pgf@x=.5\pgf@x
	}
	\anchor{south south east}{
		\pgf@process{\northeast}
		\pgf@xa=1.5\pgf@x
		\pgf@process{\southwest}
		\pgf@x=.5\pgf@x
		\advance\pgf@x by \pgf@xa
		\pgf@x=.5\pgf@x
	}
}

\pgfaddtoshape{basic}{
	\anchorlet{c}{center}		
	\anchorlet{n}{north}
	\anchorlet{e}{east}
	\anchorlet{s}{south}
	\anchorlet{w}{west}			
	\anchorlet{se}{south east}
	\anchorlet{sw}{south west}
	\anchorlet{ne}{north east}
	\anchorlet{nw}{north west}
	\anchorlet{wsw}{west south west}
	\anchorlet{wnw}{west north west}
	\anchorlet{ene}{east north east}
	\anchorlet{ese}{east south east}
	\anchorlet{nnw}{north north west}
	\anchorlet{nne}{north north east}
	\anchorlet{ssw}{south south west}
	\anchorlet{sse}{south south east}
}	

\pgfdeclareshape{ampshape}{
	\savedmacro\direction{
		\edef\direction{\pgfkeysvalueof{/tikz/ampkeys/direction}}%
	}
	\saveddimen\minwidth{
		\pgfmathsetlength\pgf@x{\pgfshapeminwidth}%
	}
	\saveddimen\minheight{
		\pgfmathsetlength\pgf@x{\pgfshapeminheight}%
	}
	\inheritsavedanchors[from=basic] 
	
	\inheritanchor[from=basic]{center}
	\inheritanchor[from=basic]{mid}		
	\inheritanchor[from=basic]{base}	
	\inheritanchor[from=basic]{north}
	\inheritanchor[from=basic]{south}		
	\inheritanchor[from=basic]{west}		
	\inheritanchor[from=basic]{mid west}				
	\inheritanchor[from=basic]{base west}		
	\inheritanchor[from=basic]{north west}		
	\inheritanchor[from=basic]{south west}		
	\inheritanchor[from=basic]{east}
	\inheritanchor[from=basic]{mid east}
	\inheritanchor[from=basic]{base east}	
	\inheritanchor[from=basic]{north east}
	\inheritanchor[from=basic]{south east}
	\inheritanchor[from=basic]{west south west}%
	\inheritanchor[from=basic]{west north west}%
	\inheritanchor[from=basic]{east north east}%
	\inheritanchor[from=basic]{east south east}%
	\inheritanchor[from=basic]{north north west}%
	\inheritanchor[from=basic]{north north east}%
	\inheritanchor[from=basic]{south south west}%
	\inheritanchor[from=basic]{south south east}%
	
	\inheritanchorborder[from=basic]	
	\inheritbackgroundpath[from=basic]

    \pgfutil@g@addto@macro\pgf@sh@s@ampshape{%
        \pgfutil@ifundefined{pgf@anchor@ampshape@in0}{
	        \expandafter\xdef\csname pgf@anchor@ampshape@in0\endcsname{%
	            \noexpand\ampshape@port{0}
	        }%
	    }{}%
        \pgfutil@ifundefined{pgf@anchor@ampshape@in}{
	        \expandafter\xdef\csname pgf@anchor@ampshape@in\endcsname{%
	            \noexpand\ampshape@port{0}
	        }%
	    }{}%
        \pgfutil@ifundefined{pgf@anchor@ampshape@out0}{
	        \expandafter\xdef\csname pgf@anchor@ampshape@out0\endcsname{%
	            \noexpand\ampshape@port{1}
	        }%
	    }{}%
        \pgfutil@ifundefined{pgf@anchor@ampshape@out}{
	        \expandafter\xdef\csname pgf@anchor@ampshape@out\endcsname{%
	            \noexpand\ampshape@port{1}
	        }%
	    }{}%
	}
}

\def\ampshape@port#1{
    \northeast	

    \ifnum#1=0	
	    \if\direction\direce
			\pgf@x=-\pgf@x
		    \pgf@ya= \pgf@y
		    \pgfmathsetlength{\pgf@y}{\pgf@ya-0.5*\minheight}%
		\fi
	    \if\direction\direcw
			\pgf@x=\pgf@x
		    \pgf@ya= \pgf@y
		    \pgfmathsetlength{\pgf@y}{\pgf@ya-0.5*\minheight}%
		\fi
	    \if\direction\direcn
			\pgf@y=-\pgf@y
		    \pgf@xa=\pgf@x
		    \pgfmathsetlength{\pgf@x}{\pgf@xa-0.5*\minwidth}%
		\fi
	    \if\direction\direcs
			\pgf@y=\pgf@y
		    \pgf@xa= \pgf@x
		    \pgfmathsetlength{\pgf@x}{\pgf@xa-0.5*\minwidth}%
		\fi
	\else	
	    \if\direction\direce
			\pgf@x=\pgf@x
		    \pgf@ya= \pgf@y
		    \pgfmathsetlength{\pgf@y}{\pgf@ya-0.5*\minheight}%
		\fi
	    \if\direction\direcw
			\pgf@x=-\pgf@x
		    \pgf@ya= \pgf@y
		    \pgfmathsetlength{\pgf@y}{\pgf@ya-0.5*\minheight}%
		\fi
	    \if\direction\direcn
			\pgf@y=\pgf@y
		    \pgf@xa= \pgf@x
		    \pgfmathsetlength{\pgf@x}{\pgf@xa-0.5*\minwidth}%
		\fi
	    \if\direction\direcs
			\pgf@y=-\pgf@y
		    \pgf@xa= \pgf@x
		    \pgfmathsetlength{\pgf@x}{\pgf@xa-0.5*\minwidth}%
		\fi
	\fi
}

\pgfaddtoshape{ampshape}{
	\anchorlet{c}{center}		
	\anchorlet{n}{north}
	\anchorlet{e}{east}
	\anchorlet{s}{south}
	\anchorlet{w}{west}			
	\anchorlet{se}{south east}
	\anchorlet{sw}{south west}
	\anchorlet{ne}{north east}
	\anchorlet{nw}{north west}
	\anchorlet{wsw}{west south west}
	\anchorlet{wnw}{west north west}
	\anchorlet{ene}{east north east}
	\anchorlet{ese}{east south east}
	\anchorlet{nnw}{north north west}
	\anchorlet{nne}{north north east}
	\anchorlet{ssw}{south south west}
	\anchorlet{sse}{south south east}
}

\tikzset{
	/tikz/ampkeys/.cd,
	height/.initial=0.5,
	width/.initial=0.5,
	color/.initial=O,
	direction/.initial=e,
	linestyle/.initial={linestyle, rounded corners = 0},
	/tikz/amp/.code={
		\pgfqkeys{/tikz/ampkeys}{#1}%
		\tikzset{/tikz/ampkeys/drawer/.expanded=%
			{\pgfkeysvalueof{/tikz/ampkeys/direction}}%
			{\pgfkeysvalueof{/tikz/ampkeys/height}}%
			{\pgfkeysvalueof{/tikz/ampkeys/width}}%
			{\pgfkeysvalueof{/tikz/ampkeys/color}}%
			{\pgfkeysvalueof{/tikz/ampkeys/linestyle}}%
		}
	},
	/tikz/ampkeys/drawer/.code n args={5}{%
		\tikzset{
			ampshape,
			minimum height=#2*\NODESIZE,
			minimum width=#3*\NODESIZE,
			append after command={
				\pgfextra{\let\bdr=\tikzlastnode%
				\if#1e
					\draw[draw=#4, fill=#4l, #5] (\bdr.sw) to (\bdr.nw) to (\bdr.e) to cycle {};
				\fi
				\if#1w
					\draw[draw=#4, fill=#4l, #5] (\bdr.se) to (\bdr.ne) to (\bdr.w) to cycle {};
				\fi
				\if#1n
					\draw[draw=#4, fill=#4l, #5] (\bdr.se) to (\bdr.sw) to (\bdr.n) to cycle {};
				\fi
				\if#1s
					\draw[draw=#4, fill=#4l, #5] (\bdr.ne) to (\bdr.nw) to (\bdr.s) to cycle {};
				\fi
				}
			}
		}
	},
}
\newcount\portcount

\pgfdeclareshape{aomshape}{
	\savedmacro\direction{
		\edef\direction{\pgfkeysvalueof{/tikz/aomkeys/direction}}%
	}
	\saveddimen\minwidth{
		\pgfmathsetlength\pgf@x{\pgfshapeminwidth}%
	}
	\saveddimen\minheight{
		\pgfmathsetlength\pgf@x{\pgfshapeminheight}%
	}
	\inheritsavedanchors[from=basic]

	\inheritanchor[from=basic]{center}
	\inheritanchor[from=basic]{mid}
	\inheritanchor[from=basic]{base}
	\inheritanchor[from=basic]{north}
	\inheritanchor[from=basic]{south}
	\inheritanchor[from=basic]{west}
	\inheritanchor[from=basic]{mid west}
	\inheritanchor[from=basic]{base west}
	\inheritanchor[from=basic]{north west}
	\inheritanchor[from=basic]{south west}
	\inheritanchor[from=basic]{east}
	\inheritanchor[from=basic]{mid east}
	\inheritanchor[from=basic]{base east}
	\inheritanchor[from=basic]{north east}
	\inheritanchor[from=basic]{south east}
	\inheritanchor[from=basic]{west south west}%
	\inheritanchor[from=basic]{west north west}%
	\inheritanchor[from=basic]{east north east}%
	\inheritanchor[from=basic]{east south east}%
	\inheritanchor[from=basic]{north north west}%
	\inheritanchor[from=basic]{north north east}%
	\inheritanchor[from=basic]{south south west}%
	\inheritanchor[from=basic]{south south east}%

	\inheritanchorborder[from=basic]
	\inheritbackgroundpath[from=basic]

	\pgfutil@g@addto@macro\pgf@sh@s@aomshape{%
		\pgfutil@ifundefined{pgf@anchor@aomshape@in0}{
			\expandafter\xdef\csname pgf@anchor@aomshape@in0\endcsname{%
				\noexpand\aomshape@port{0}
			}%
		}{}%
		\pgfutil@ifundefined{pgf@anchor@aomshape@in}{
			\expandafter\xdef\csname pgf@anchor@aomshape@in\endcsname{%
				\noexpand\aomshape@port{0}
			}%
		}{}%
		\pgfutil@ifundefined{pgf@anchor@aomshape@out0}{
			\expandafter\xdef\csname pgf@anchor@aomshape@out0\endcsname{%
				\noexpand\aomshape@port{1}
			}%
		}{}%
		\pgfutil@ifundefined{pgf@anchor@aomshape@out}{
			\expandafter\xdef\csname pgf@anchor@aomshape@out\endcsname{%
				\noexpand\aomshape@port{1}
			}%
		}{}%
	}
}

\def\aomshape@port#1{
	\northeast	

	\ifnum#1=0	
		\if\direction\direce
			\pgf@x=-\pgf@x
			\pgf@ya= \pgf@y
			\pgfmathsetlength{\pgf@y}{\pgf@ya-0.5*\minheight}%
		\fi
		\if\direction\direcw
			\pgf@x=\pgf@x
			\pgf@ya= \pgf@y
			\pgfmathsetlength{\pgf@y}{\pgf@ya-0.5*\minheight}%
		\fi
		\if\direction\direcn
			\pgf@y=-\pgf@y
			\pgf@xa=\pgf@x
			\pgfmathsetlength{\pgf@x}{\pgf@xa-0.5*\minwidth}%
		\fi
		\if\direction\direcs
			\pgf@y=\pgf@y
			\pgf@xa= \pgf@x
			\pgfmathsetlength{\pgf@x}{\pgf@xa-0.5*\minwidth}%
		\fi
	\else	
		\if\direction\direce
			\pgf@x=\pgf@x
			\pgf@ya= \pgf@y
			\pgfmathsetlength{\pgf@y}{\pgf@ya-0.5*\minheight}%
		\fi
		\if\direction\direcw
			\pgf@x=-\pgf@x
			\pgf@ya= \pgf@y
			\pgfmathsetlength{\pgf@y}{\pgf@ya-0.5*\minheight}%
		\fi
		\if\direction\direcn
			\pgf@y=\pgf@y
			\pgf@xa= \pgf@x
			\pgfmathsetlength{\pgf@x}{\pgf@xa-0.5*\minwidth}%
		\fi
		\if\direction\direcs
			\pgf@y=-\pgf@y
			\pgf@xa= \pgf@x
			\pgfmathsetlength{\pgf@x}{\pgf@xa-0.5*\minwidth}%
		\fi
	\fi
}

\pgfaddtoshape{aomshape}{
	\anchorlet{c}{center}
	\anchorlet{n}{north}
	\anchorlet{e}{east}
	\anchorlet{s}{south}
	\anchorlet{w}{west}
	\anchorlet{se}{south east}
	\anchorlet{sw}{south west}
	\anchorlet{ne}{north east}
	\anchorlet{nw}{north west}
	\anchorlet{wsw}{west south west}
	\anchorlet{wnw}{west north west}
	\anchorlet{ene}{east north east}
	\anchorlet{ese}{east south east}
	\anchorlet{nnw}{north north west}
	\anchorlet{nne}{north north east}
	\anchorlet{ssw}{south south west}
	\anchorlet{sse}{south south east}
}

\tikzset{
/tikz/aomkeys/.cd,
size/.initial=1,
circlesize/.initial=1,
color/.initial=O,
direction/.initial=e,
linestyle/.initial={linestyle, inner sep=0.5mm},
fillgradient/.initial=O,
/tikz/aom/.code={
\pgfqkeys{/tikz/aomkeys}{#1}%
\tikzset{/tikz/aomkeys/drawer/.expanded=%
	{\pgfkeysvalueof{/tikz/aomkeys/size}}%
	{\pgfkeysvalueof{/tikz/aomkeys/color}}%
	{\pgfkeysvalueof{/tikz/aomkeys/linestyle}}%
\if\pgfkeysvalueof{/tikz/aomkeys/direction}e
	{0}%
\fi
\if\pgfkeysvalueof{/tikz/aomkeys/direction}w
	{0}%
\fi
\if\pgfkeysvalueof{/tikz/aomkeys/direction}n
	{1}%
\fi
\if\pgfkeysvalueof{/tikz/aomkeys/direction}s
	{1}%
\fi
{\pgfkeysvalueof{/tikz/aomkeys/direction}}%
{\pgfkeysvalueof{/tikz/aomkeys/circlesize}}%
{\pgfkeysvalueof{/tikz/aomkeys/fillgradient}}%
}
},
/tikz/aomkeys/drawer/.code n args={7}{%
		\tikzset{
			aomshape,
			draw,
			minimum height = #1*\NODESIZE,
			minimum width = #1*\NODESIZE,
			#2,
			#3,
			append after command={
					\pgfextra{\let\bdr=\tikzlastnode%
						\node[#7, fit=(\bdr.nw)(\bdr.se)] (boxgradient){};

						\node[coordinate] at ($(\bdr.in)!0.25!(\bdr.out)$) (circlein){};
						\node[coordinate] at ($(\bdr.in)!0.75!(\bdr.out)$) (circleout){};

						\ifnum#4>0
							\node[coordinate] at (circleout -| \bdr.nne) (circleouttop){};
						\else
							\node[coordinate] at (circleout |- \bdr.ene) (circleouttop){};
						\fi

						\draw[---, #2, #3, fill] (\bdr.in) to (circlein) circle (0.05*#6);
						\draw[---, #2, #3, fill] (\bdr.out) to (circleout) circle (0.05*#6);

						\draw[---, #2, #3] (circlein) to (circleouttop){};

					}
				}
		}
	},
}
\newcount\portcount

\pgfdeclareshape{boxshape}{
	\savedmacro\nin{
		\edef\nin{\pgfkeysvalueof{/tikz/boxkeys/nin}}%
	}
	\savedmacro\nout{
		\edef\nout{\pgfkeysvalueof{/tikz/boxkeys/nout}}%
	}
	\savedmacro\direction{
		\edef\direction{\pgfkeysvalueof{/tikz/boxkeys/direction}}%
	}
	\inheritsavedanchors[from=basic]

	\inheritanchor[from=basic]{center}
	\inheritanchor[from=basic]{mid}
	\inheritanchor[from=basic]{base}
	\inheritanchor[from=basic]{north}
	\inheritanchor[from=basic]{south}
	\inheritanchor[from=basic]{west}
	\inheritanchor[from=basic]{mid west}
	\inheritanchor[from=basic]{base west}
	\inheritanchor[from=basic]{north west}
	\inheritanchor[from=basic]{south west}
	\inheritanchor[from=basic]{east}
	\inheritanchor[from=basic]{mid east}
	\inheritanchor[from=basic]{base east}
	\inheritanchor[from=basic]{north east}
	\inheritanchor[from=basic]{south east}
	\inheritanchor[from=basic]{west south west}%
	\inheritanchor[from=basic]{west north west}%
	\inheritanchor[from=basic]{east north east}%
	\inheritanchor[from=basic]{east south east}%
	\inheritanchor[from=basic]{north north west}%
	\inheritanchor[from=basic]{north north east}%
	\inheritanchor[from=basic]{south south west}%
	\inheritanchor[from=basic]{south south east}%

	\inheritanchorborder[from=basic]
	\inheritbackgroundpath[from=basic]

	\pgfutil@g@addto@macro\pgf@sh@s@boxshape{%
		\pgfmathsetcount{\portcount}{0}
		\pgfmathloop%
		\ifnum\the\portcount<\nin
		\pgfutil@ifundefined{pgf@anchor@boxshape@in\the\portcount}{
			\expandafter\xdef\csname pgf@anchor@boxshape@in\the\portcount\endcsname{%
				\noexpand\boxshape@port[\the\portcount]{0}
			}%
		}{}%
		\ifnum\the\portcount=0
			\pgfutil@ifundefined{pgf@anchor@boxshape@in}{%
				\expandafter\xdef\csname pgf@anchor@boxshape@in\endcsname{%
					\noexpand\boxshape@port[\the\portcount]{0}
				}%
			}{}%
		\fi
		\pgfmathaddtocount{\portcount}{1}	
		\repeatpgfmathloop
		\pgfmathsetcount{\portcount}{0}
		\pgfmathloop%
		\ifnum\the\portcount<\nout
		\pgfutil@ifundefined{pgf@anchor@boxshape@out\the\portcount}{%
			\expandafter\xdef\csname pgf@anchor@boxshape@out\the\portcount\endcsname{%
				\noexpand\boxshape@port[\the\portcount]{1}
			}%
		}{}%
		\ifnum\the\portcount=0
			\pgfutil@ifundefined{pgf@anchor@boxshape@out}{%
				\expandafter\xdef\csname pgf@anchor@boxshape@out\endcsname{%
					\noexpand\boxshape@port[\the\portcount]{1}
				}%
			}{}%
		\fi
		\pgfmathaddtocount{\portcount}{1}	
		\repeatpgfmathloop
	}
}

\def\boxshape@port[#1]#2{
	\northeast \pgf@xa=\pgf@x \pgf@ya=\pgf@y
	\southwest \pgf@xb=\pgf@x \pgf@yb=\pgf@y

	\ifnum#2=0	
		\if\direction\direce	
			\pgf@x=\pgf@xb
			\pgf@yc=\pgf@ya \advance\pgf@yc by -\pgf@yb	
			\pgfmathsetlength{\pgf@y}{\pgf@ya-(#1 + 0.5)*(\pgf@yc/\nin)}%
		\fi
		\if\direction\direcw
			\pgf@x=\pgf@xa
			\pgf@yc=\pgf@ya \advance\pgf@yc by -\pgf@yb	
			\pgfmathsetlength{\pgf@y}{\pgf@ya-(#1 + 0.5)*(\pgf@yc/\nin)}%
		\fi
		\if\direction\direcn
			\pgf@y=\pgf@yb
			\pgf@xc=\pgf@xa \advance\pgf@xc by -\pgf@xb	
			\pgfmathsetlength{\pgf@x}{\pgf@xb+(#1 + 0.5)*(\pgf@xc/\nin)}%
		\fi
		\if\direction\direcs
			\pgf@y=\pgf@ya
			\pgf@xc=\pgf@xa \advance\pgf@xc by -\pgf@xb	
			\pgfmathsetlength{\pgf@x}{\pgf@xb+(#1 + 0.5)*(\pgf@xc/\nin)}%
		\fi
	\else	
		\if\direction\direce	
			\pgf@x=\pgf@xa
			\pgf@yc=\pgf@ya \advance\pgf@yc by -\pgf@yb	
			\pgfmathsetlength{\pgf@y}{\pgf@ya-(#1 + 0.5)*(\pgf@yc/\nout)}%
		\fi
		\if\direction\direcw
			\pgf@x=\pgf@xb
			\pgf@yc=\pgf@ya \advance\pgf@yc by -\pgf@yb	
			\pgfmathsetlength{\pgf@y}{\pgf@ya-(#1 + 0.5)*(\pgf@yc/\nout)}%
		\fi
		\if\direction\direcn
			\pgf@y=\pgf@ya
			\pgf@xc=\pgf@xa \advance\pgf@xc by -\pgf@xb	
			\pgfmathsetlength{\pgf@x}{\pgf@xb+(#1 + 0.5)*(\pgf@xc/\nout)}%
		\fi
		\if\direction\direcs
			\pgf@y=\pgf@yb
			\pgf@xc=\pgf@xa \advance\pgf@xc by -\pgf@xb	
			\pgfmathsetlength{\pgf@x}{\pgf@xb+(#1 + 0.5)*(\pgf@xc/\nout)}%
		\fi
	\fi
}

\pgfaddtoshape{boxshape}{
	\anchorlet{c}{center}
	\anchorlet{n}{north}
	\anchorlet{e}{east}
	\anchorlet{s}{south}
	\anchorlet{w}{west}
	\anchorlet{se}{south east}
	\anchorlet{sw}{south west}
	\anchorlet{ne}{north east}
	\anchorlet{nw}{north west}
	\anchorlet{wsw}{west south west}
	\anchorlet{wnw}{west north west}
	\anchorlet{ene}{east north east}
	\anchorlet{ese}{east south east}
	\anchorlet{nnw}{north north west}
	\anchorlet{nne}{north north east}
	\anchorlet{ssw}{south south west}
	\anchorlet{sse}{south south east}
}

\tikzset{
/tikz/boxkeys/.cd,
height/.initial=0.5,
width/.initial=1,
color/.initial=O,
direction/.initial=e,
linestyle/.initial={linestyle, inner sep=0.5mm},
nin/.initial=1,
nout/.initial=1,
draw/.initial=1,
/tikz/box/.code={
\pgfqkeys{/tikz/boxkeys}{#1}%
\tikzset{/tikz/boxkeys/drawer/.expanded=%
\if\pgfkeysvalueof{/tikz/boxkeys/direction}e
	{\pgfkeysvalueof{/tikz/boxkeys/width}}%
	{\pgfkeysvalueof{/tikz/boxkeys/height}}%
	{0}%
	{-90}%
\fi
\if\pgfkeysvalueof{/tikz/boxkeys/direction}w
	{\pgfkeysvalueof{/tikz/boxkeys/width}}%
	{\pgfkeysvalueof{/tikz/boxkeys/height}}%
	{0}%
	{90}%
\fi
\if\pgfkeysvalueof{/tikz/boxkeys/direction}n
	{\pgfkeysvalueof{/tikz/boxkeys/width}}%
	{\pgfkeysvalueof{/tikz/boxkeys/height}}%
	{1}%
	{0}%
\fi
\if\pgfkeysvalueof{/tikz/boxkeys/direction}s
	{\pgfkeysvalueof{/tikz/boxkeys/width}}%
	{\pgfkeysvalueof{/tikz/boxkeys/height}}%
	{1}%
	{180}%
\fi
{\pgfkeysvalueof{/tikz/boxkeys/color}}%
{\pgfkeysvalueof{/tikz/boxkeys/linestyle}}%
\ifnum\pgfkeysvalueof{/tikz/boxkeys/draw}>0%
	{draw}%
\else
	{}
\fi
}
},
/tikz/boxkeys/drawer/.code n args={7}{%
		\tikzset{
			boxshape,
			#7,
			#6,
			#5,
			minimum height=
			\ifnum#3>0	
				#1*\NODESIZE
			\else
				#2*\NODESIZE
			\fi
			,minimum width=
			\ifnum#3>0
				#2*\NODESIZE
			\else
				#1*\NODESIZE
			\fi
		}
	},
}
\pgfdeclareshape{beshape}{ 
    \savedmacro\nports{
        \edef\nports{\pgfkeysvalueof{/tikz/bekeys/nports}}%
    }
    \savedmacro\direction{
        \edef\direction{\pgfkeysvalueof{/tikz/bekeys/direction}}%
    }
    \savedmacro\inverted{
        \edef\inverted{\pgfkeysvalueof{/tikz/bekeys/inverted}}%
    }
    \savedmacro\ninports{
        \ifnum\inverted=0
        \edef\ninports{\nports}
        \else
        \edef\ninports{1}%
        \fi
    }
    \savedmacro\noutports{
        \ifnum\inverted=0
        \edef\noutports{1}%
        \else
        \edef\noutports{\nports}%
        \fi
    }
    \inheritsavedanchors[from=basic]

    \inheritanchor[from=basic]{center}
    \inheritanchor[from=basic]{mid}
    \inheritanchor[from=basic]{base}
    \inheritanchor[from=basic]{north}
    \inheritanchor[from=basic]{south}
    \inheritanchor[from=basic]{west}
    \inheritanchor[from=basic]{mid west}
    \inheritanchor[from=basic]{base west}
    \inheritanchor[from=basic]{north west}
    \inheritanchor[from=basic]{south west}
    \inheritanchor[from=basic]{east}
    \inheritanchor[from=basic]{mid east}
    \inheritanchor[from=basic]{base east}
    \inheritanchor[from=basic]{north east}
    \inheritanchor[from=basic]{south east}
    \inheritanchor[from=basic]{west south west}%
    \inheritanchor[from=basic]{west north west}%
    \inheritanchor[from=basic]{east north east}%
    \inheritanchor[from=basic]{east south east}%
    \inheritanchor[from=basic]{north north west}%
    \inheritanchor[from=basic]{north north east}%
    \inheritanchor[from=basic]{south south west}%
    \inheritanchor[from=basic]{south south east}%

    \inheritanchorborder[from=basic]
    \inheritbackgroundpath[from=basic]

    \pgfutil@g@addto@macro\pgf@sh@s@beshape{%
        \pgfmathsetcount{\portcount}{0}
        \pgfmathloop%
        \ifnum\the\portcount<\nports
        \ifnum\the\portcount<\ninports
        \pgfutil@ifundefined{pgf@anchor@beshape@in\the\portcount}{
            \expandafter\xdef\csname pgf@anchor@beshape@in\the\portcount\endcsname{%
                \noexpand\beshape@port[\the\portcount]{0}
            }%
        }{}%
        \ifnum\the\portcount=0
        \pgfutil@ifundefined{pgf@anchor@beshape@in}{%
            \expandafter\xdef\csname pgf@anchor@beshape@in\endcsname{%
                \noexpand\beshape@port[\the\portcount]{0}
            }%
        }{}%
        \fi
        \fi
        \ifnum\the\portcount<\noutports
        \pgfutil@ifundefined{pgf@anchor@beshape@out\the\portcount}{%
            \expandafter\xdef\csname pgf@anchor@beshape@out\the\portcount\endcsname{%
                \noexpand\beshape@port[\the\portcount]{1}
            }%
        }{}%
        \ifnum\the\portcount=0
        \pgfutil@ifundefined{pgf@anchor@beshape@out}{%
            \expandafter\xdef\csname pgf@anchor@beshape@out\endcsname{%
                \noexpand\beshape@port[\the\portcount]{1}
            }%
        }{}%
        \fi
        \fi
        \pgfmathaddtocount{\portcount}{1}    
        \repeatpgfmathloop
    }
}

\def\beshape@port[#1]#2{
    \northeast \pgf@xa=\pgf@x \pgf@ya=\pgf@y
    \southwest \pgf@xb=\pgf@x \pgf@yb=\pgf@y

    \ifnum#2=0
    \ifnum\inverted=0
    \def\chooseports{0}
    \else
    \def\chooseports{1}
    \fi
    \else
    \ifnum\inverted=0
    \def\chooseports{1}
    \else
    \def\chooseports{0}
    \fi
    \fi

    \ifnum\chooseports=0    
    \if\direction\direce
    \pgf@x=\pgf@xb
    \pgf@yc=\pgf@ya \advance\pgf@yc by -\pgf@yb    
    \pgfmathsetlength{\pgf@y}{\pgf@ya-(#1 + 0.5)*(\pgf@yc/\nports)}%
    \fi
    \if\direction\direcw
    \pgf@x=\pgf@xa
    \pgf@yc=\pgf@ya \advance\pgf@yc by -\pgf@yb    
    \pgfmathsetlength{\pgf@y}{\pgf@ya-(#1 + 0.5)*(\pgf@yc/\nports)}%
    \fi
    \if\direction\direcn
    \pgf@y=\pgf@yb
    \pgf@xc=\pgf@xa \advance\pgf@xc by -\pgf@xb    
    \pgfmathsetlength{\pgf@x}{\pgf@xb+(#1 + 0.5)*(\pgf@xc/\nports)}%
    \fi
    \if\direction\direcs
    \pgf@y=\pgf@ya
    \pgf@xc=\pgf@xa \advance\pgf@xc by -\pgf@xb    
    \pgfmathsetlength{\pgf@x}{\pgf@xb+(#1 + 0.5)*(\pgf@xc/\nports)}%
    \fi
    \else    
    \if\direction\direce
    \pgf@x=\pgf@xa
    \pgf@yc=\pgf@ya \advance\pgf@yc by -\pgf@yb    
    \pgfmathsetlength{\pgf@y}{\pgf@ya-0.5\pgf@yc}%
    \fi
    \if\direction\direcw
    \pgf@x=\pgf@xb
    \pgf@yc=\pgf@ya \advance\pgf@yc by -\pgf@yb    
    \pgfmathsetlength{\pgf@y}{\pgf@ya-0.5\pgf@yc}%
    \fi
    \if\direction\direcn
    \pgf@y=\pgf@ya
    \pgf@xc=\pgf@xa \advance\pgf@xc by -\pgf@xb    
    \pgfmathsetlength{\pgf@x}{\pgf@xa-0.5\pgf@xc}%
    \fi
    \if\direction\direcs
    \pgf@y=\pgf@yb
    \pgf@xc=\pgf@xa \advance\pgf@xc by -\pgf@xb    
    \pgfmathsetlength{\pgf@x}{\pgf@xa-0.5\pgf@xc}%
    \fi
    \fi
}

\pgfaddtoshape{beshape}{
    \anchorlet{c}{center}
    \anchorlet{n}{north}
    \anchorlet{e}{east}
    \anchorlet{s}{south}
    \anchorlet{w}{west}
    \anchorlet{se}{south east}
    \anchorlet{sw}{south west}
    \anchorlet{ne}{north east}
    \anchorlet{nw}{north west}
    \anchorlet{wsw}{west south west}
    \anchorlet{wnw}{west north west}
    \anchorlet{ene}{east north east}
    \anchorlet{ese}{east south east}
    \anchorlet{nnw}{north north west}
    \anchorlet{nne}{north north east}
    \anchorlet{ssw}{south south west}
    \anchorlet{sse}{south south east}
}

\tikzset{
    /tikz/bekeys/.cd,
    height/.initial=1,
    width/.initial=0.5,
    color/.initial=O,
    direction/.initial=e,
    linestyle/.initial={linestyle, rounded corners = 0},
    nports/.initial=3,
    inverted/.initial=0,    
    sbe/.initial=0,
    angle/.initial=60,
    /tikz/be/.code={
        \pgfqkeys{/tikz/bekeys}{#1}%
        \tikzset{/tikz/bekeys/drawer/.expanded=%
            \if\pgfkeysvalueof{/tikz/bekeys/direction}e
                {\pgfkeysvalueof{/tikz/bekeys/width}}%
                {\pgfkeysvalueof{/tikz/bekeys/height}}%
                {0}%
                {-90}%
            \fi
            \if\pgfkeysvalueof{/tikz/bekeys/direction}w
                {\pgfkeysvalueof{/tikz/bekeys/width}}%
                {\pgfkeysvalueof{/tikz/bekeys/height}}%
                {0}%
                {90}%
            \fi
            \if\pgfkeysvalueof{/tikz/bekeys/direction}n
                {\pgfkeysvalueof{/tikz/bekeys/width}}%
                {\pgfkeysvalueof{/tikz/bekeys/height}}%
                {1}%
                {0}%
            \fi
            \if\pgfkeysvalueof{/tikz/bekeys/direction}s
                {\pgfkeysvalueof{/tikz/bekeys/width}}%
                {\pgfkeysvalueof{/tikz/bekeys/height}}%
                {1}%
                {180}%
            \fi
            {\pgfkeysvalueof{/tikz/bekeys/color}}%
            {\pgfkeysvalueof{/tikz/bekeys/linestyle}}%
            {\pgfkeysvalueof{/tikz/bekeys/sbe}}%
            {\pgfkeysvalueof{/tikz/bekeys/angle}}%
        }
    },
    /tikz/bekeys/drawer/.code n args={8}{%
        \tikzset{
            beshape,
            #6,
            #5,
            minimum height=
            \ifnum#3>0    
            #1*\NODESIZE
            \else
            #2*\NODESIZE
            \fi
            ,minimum width=
            \ifnum#3>0
            #2*\NODESIZE
            \else
            #1*\NODESIZE
            \fi
            ,append after command={
                \pgfextra{
                    \let\bdr=\tikzlastnode%
                    \node[trapezium, line width = \NODETHICKNESS, minimum height=#1*\NODESIZE, minimum width=#2*\NODESIZE, trapezium stretches=true, rotate=#4, trapezium angle=70, inner sep=0.001mm, #5, #6] at (\bdr) (trap) {};

                    \node[rectangle, line width = \NODETHICKNESS, minimum height=#1*\NODESIZE, minimum width=#2*\NODESIZE, anchor=north, rotate=#4, #5, #6] at (trap.south) (r1) {};

                    \tikzmath{coordinate \C;
                    \C = (trap.top left corner)-(trap.top right corner);
                    \distAB = sqrt((\Cx)^2+(\Cy)^2);
                    }

                    \node[rectangle, line width = \NODETHICKNESS, minimum height=#1*\NODESIZE, minimum width=\distAB, anchor=south, rotate=#4, #5, #6, red] at (trap.north) (r2) {};

					\draw[#5, #6] (trap.top left corner) to (r2.north west) to (r2.north east) to (trap.top right corner) to (trap.bottom right corner) to (r1.south east) to (r1.south west) to (r1.north west) to cycle;

                }
            }
        }
    },
}
\pgfdeclareshape{couplershape}{	

	\inheritsavedanchors[from=basic] 

	\inheritanchor[from=basic]{center}
	\inheritanchor[from=basic]{mid}		
	\inheritanchor[from=basic]{base}	
	\inheritanchor[from=basic]{north}
	\inheritanchor[from=basic]{south}		
	\inheritanchor[from=basic]{west}		
	\inheritanchor[from=basic]{mid west}				
	\inheritanchor[from=basic]{base west}		
	\inheritanchor[from=basic]{north west}		
	\inheritanchor[from=basic]{south west}		
	\inheritanchor[from=basic]{east}
	\inheritanchor[from=basic]{mid east}
	\inheritanchor[from=basic]{base east}	
	\inheritanchor[from=basic]{north east}
	\inheritanchor[from=basic]{south east}
	\inheritanchor[from=basic]{west south west}%
	\inheritanchor[from=basic]{west north west}%
	\inheritanchor[from=basic]{east north east}%
	\inheritanchor[from=basic]{east south east}%
	\inheritanchor[from=basic]{north north west}%
	\inheritanchor[from=basic]{north north east}%
	\inheritanchor[from=basic]{south south west}%
	\inheritanchor[from=basic]{south south east}%
	
	\inheritanchorborder[from=basic]	
	\inheritbackgroundpath[from=basic]
}

\pgfaddtoshape{couplershape}{
	\anchorlet{in}{center}		
	\anchorlet{in0}{center}
	\anchorlet{out}{center}
	\anchorlet{out0}{center}
	\anchorlet{c}{center}		
	\anchorlet{n}{north}
	\anchorlet{e}{east}
	\anchorlet{s}{south}
	\anchorlet{w}{west}			
	\anchorlet{se}{south east}
	\anchorlet{sw}{south west}
	\anchorlet{ne}{north east}
	\anchorlet{nw}{north west}
	\anchorlet{wsw}{west south west}
	\anchorlet{wnw}{west north west}
	\anchorlet{ene}{east north east}
	\anchorlet{ese}{east south east}
	\anchorlet{nnw}{north north west}
	\anchorlet{nne}{north north east}
	\anchorlet{ssw}{south south west}
	\anchorlet{sse}{south south east}
}

\tikzset{
	/tikz/couplerkeys/.cd,
	size/.initial=0.2,
	color/.initial=O,
	rotation/.initial=0,
	heightwidthratio/.initial=0.5,
	/tikz/coupler/.code={
		\pgfqkeys{/tikz/couplerkeys}{#1}%
		\tikzset{/tikz/couplerkeys/drawer/.expanded=%
			{\pgfkeysvalueof{/tikz/couplerkeys/size}}%
			{\pgfkeysvalueof{/tikz/couplerkeys/color}}%
			{\pgfkeysvalueof{/tikz/couplerkeys/rotation}}%
			{\pgfkeysvalueof{/tikz/couplerkeys/heightwidthratio}}%
		}
	},
	/tikz/couplerkeys/drawer/.code n args={4}{%
		\tikzset{
			couplershape,
			minimum height=#1*\NODESIZE
			\ifnum#3<1
				\ifnum#3>-1
					*#4
				\fi
			\fi
			,minimum width=#1*\NODESIZE
			\ifnum#3<91
				\ifnum#3>89
					*#4
				\fi
			\fi
			\ifnum#3<-89
				\ifnum#3>-91
					*#4
				\fi
			\fi
			,#2,
			append after command={
				\pgfextra{\let\bdr=\tikzlastnode%
				\node[ellipse, fill, #2, rotate=#3, outer sep = 0, minimum width=#1*\NODESIZE, minimum height=#1*#4*\NODESIZE] at (\bdr.center){};
				}
			}
		}
	},
}
\pgfdeclareshape{fibershape}{
	\savedmacro\direction{
		\edef\direction{\pgfkeysvalueof{/tikz/fiberkeys/direction}}%
	}
	\savedmacro\flip{
		\edef\flip{\pgfkeysvalueof{/tikz/fiberkeys/flip}}%
	}
	\saveddimen\minwidth{
		\pgfmathsetlength\pgf@x{\pgfshapeminwidth}%
	}
	\saveddimen\minheight{
		\pgfmathsetlength\pgf@x{\pgfshapeminheight}%
	}
	\inheritsavedanchors[from=basic] 
	
	\inheritanchor[from=basic]{center}
	\inheritanchor[from=basic]{mid}		
	\inheritanchor[from=basic]{base}	
	\inheritanchor[from=basic]{north}
	\inheritanchor[from=basic]{south}		
	\inheritanchor[from=basic]{west}		
	\inheritanchor[from=basic]{mid west}				
	\inheritanchor[from=basic]{base west}		
	\inheritanchor[from=basic]{north west}		
	\inheritanchor[from=basic]{south west}		
	\inheritanchor[from=basic]{east}
	\inheritanchor[from=basic]{mid east}
	\inheritanchor[from=basic]{base east}	
	\inheritanchor[from=basic]{north east}
	\inheritanchor[from=basic]{south east}
	\inheritanchor[from=basic]{west south west}%
	\inheritanchor[from=basic]{west north west}%
	\inheritanchor[from=basic]{east north east}%
	\inheritanchor[from=basic]{east south east}%
	\inheritanchor[from=basic]{north north west}%
	\inheritanchor[from=basic]{north north east}%
	\inheritanchor[from=basic]{south south west}%
	\inheritanchor[from=basic]{south south east}%
	
	\inheritanchorborder[from=basic]	
	\inheritbackgroundpath[from=basic]

    \pgfutil@g@addto@macro\pgf@sh@s@fibershape{%
        \pgfutil@ifundefined{pgf@anchor@fibershape@in0}{
	        \expandafter\xdef\csname pgf@anchor@fibershape@in0\endcsname{%
	            \noexpand\fibershape@port{0}
	        }%
	    }{}%
        \pgfutil@ifundefined{pgf@anchor@fibershape@in}{
	        \expandafter\xdef\csname pgf@anchor@fibershape@in\endcsname{%
	            \noexpand\fibershape@port{0}
	        }%
	    }{}%
        \pgfutil@ifundefined{pgf@anchor@fibershape@out0}{
	        \expandafter\xdef\csname pgf@anchor@fibershape@out0\endcsname{%
	            \noexpand\fibershape@port{1}
	        }%
	    }{}%
        \pgfutil@ifundefined{pgf@anchor@fibershape@out}{
	        \expandafter\xdef\csname pgf@anchor@fibershape@out\endcsname{%
	            \noexpand\fibershape@port{1}
	        }%
	    }{}%
	}
}

\def\fibershape@port#1{
    \northeast	

    \ifnum#1=0	
	    \if\direction\direce
			\pgf@x=-\pgf@x
	    	\if\flip\flipfalse
		    	\pgf@y=-\pgf@y
		    \else
		    	\pgf@y=\pgf@y
		    \fi
		\fi
	    \if\direction\direcw
			\pgf@x=\pgf@x
	    	\if\flip\flipfalse
		    	\pgf@y=-\pgf@y
		    \else
		    	\pgf@y=\pgf@y
		    \fi
		\fi
	    \if\direction\direcn
			\pgf@y=-\pgf@y
	    	\if\flip\flipfalse
		    	\pgf@x=-\pgf@x
		    \else
		    	\pgf@x=\pgf@x
		    \fi
		\fi
	    \if\direction\direcs
			\pgf@y=\pgf@y
	    	\if\flip\flipfalse
		    	\pgf@x=-\pgf@x
		    \else
		    	\pgf@x=\pgf@x
		    \fi
		\fi
	\else	
	    \if\direction\direce
			\pgf@x=\pgf@x
	    	\if\flip\flipfalse
		    	\pgf@y=-\pgf@y
		    \else
		    	\pgf@y=\pgf@y
		    \fi
		\fi
	    \if\direction\direcw
			\pgf@x=-\pgf@x
	    	\if\flip\flipfalse
		    	\pgf@y=-\pgf@y
		    \else
		    	\pgf@y=\pgf@y
		    \fi
		\fi
	    \if\direction\direcn
			\pgf@y=\pgf@y
	    	\if\flip\flipfalse
		    	\pgf@x=-\pgf@x
		    \else
		    	\pgf@x=\pgf@x
		    \fi
		\fi
	    \if\direction\direcs
			\pgf@y=-\pgf@y
	    	\if\flip\flipfalse
		    	\pgf@x=-\pgf@x
		    \else
		    	\pgf@x=\pgf@x
		    \fi
		\fi
	\fi
}

\pgfaddtoshape{fibershape}{
	\anchorlet{c}{center}		
	\anchorlet{n}{north}
	\anchorlet{e}{east}
	\anchorlet{s}{south}
	\anchorlet{w}{west}			
	\anchorlet{se}{south east}
	\anchorlet{sw}{south west}
	\anchorlet{ne}{north east}
	\anchorlet{nw}{north west}
	\anchorlet{wsw}{west south west}
	\anchorlet{wnw}{west north west}
	\anchorlet{ene}{east north east}
	\anchorlet{ese}{east south east}
	\anchorlet{nnw}{north north west}
	\anchorlet{nne}{north north east}
	\anchorlet{ssw}{south south west}
	\anchorlet{sse}{south south east}
}

\tikzset{
	/tikz/fiberkeys/.cd,
	size/.initial=1,
	color/.initial=C0,
	direction/.initial=e,
	linestyle/.initial={linestyle},
	flip/.initial={0},
	drawbase/.initial={1},
	/tikz/fiber/.code={
		\pgfqkeys{/tikz/fiberkeys}{#1}%
		\tikzset{/tikz/fiberkeys/drawer/.expanded=%
			{\pgfkeysvalueof{/tikz/fiberkeys/direction}}%
			{\pgfkeysvalueof{/tikz/fiberkeys/size}}%
			{\pgfkeysvalueof{/tikz/fiberkeys/color}}%
			{\pgfkeysvalueof{/tikz/fiberkeys/linestyle}}%
			\if\pgfkeysvalueof{/tikz/fiberkeys/direction}e
				{a}%
				\if\pgfkeysvalueof{/tikz/fiberkeys/flip}0
					{south}%
				\else
					{north}%
				\fi
				{\pgfkeysvalueof{/tikz/fiberkeys/size}}
				{\pgfkeysvalueof{/tikz/fiberkeys/size} * 0.5}
			\fi
			\if\pgfkeysvalueof{/tikz/fiberkeys/direction}w
				{a}%
				\if\pgfkeysvalueof{/tikz/fiberkeys/flip}0
					{south}%
				\else
					{north}%
				\fi
				{\pgfkeysvalueof{/tikz/fiberkeys/size}}
				{\pgfkeysvalueof{/tikz/fiberkeys/size} * 0.5}
			\fi
			\if\pgfkeysvalueof{/tikz/fiberkeys/direction}n
				{b}%
				\if\pgfkeysvalueof{/tikz/fiberkeys/flip}0
					{west}%
				\else
					{east}%
				\fi
				{\pgfkeysvalueof{/tikz/fiberkeys/size} * 0.5}
				{\pgfkeysvalueof{/tikz/fiberkeys/size}}
			\fi
			\if\pgfkeysvalueof{/tikz/fiberkeys/direction}s
				{b}%
				\if\pgfkeysvalueof{/tikz/fiberkeys/flip}0
					{west}%
				\else
					{east}%
				\fi
				{\pgfkeysvalueof{/tikz/fiberkeys/size} * 0.5}
				{\pgfkeysvalueof{/tikz/fiberkeys/size}}
			\fi
			{\pgfkeysvalueof{/tikz/fiberkeys/drawbase}}%
		}
	},
	/tikz/fiberkeys/drawer/.code n args={9}{%
		\tikzset{
			fibershape,
			minimum width=#7*\NODESIZE,
			minimum height=#8*\NODESIZE,
			append after command={
				\pgfextra{\let\bdr=\tikzlastnode%
				\if#5a	
					\ifnum#9>0
						\draw[#3, #4] (\bdr.#6 west) to (\bdr.#6 east) {};
					\fi
					\node[draw=#3, #4, circle, minimum size=#2*0.5*\NODESIZE, anchor=#6] at ([xshift=-0.1*#2*\NODESIZE]\bdr.#6) () {};
					\node[draw=#3, #4, circle, minimum size=#2*0.5*\NODESIZE, anchor=#6] at (\bdr.#6) () {};
					\node[draw=#3, #4, circle, minimum size=#2*0.5*\NODESIZE, anchor=#6] at ([xshift=0.1*#2*\NODESIZE]\bdr.#6) () {};
				\fi
				\if#5b	
					\ifnum#9>0
						\draw[#3, #4] (\bdr.north #6) to (\bdr.south #6) {};
					\fi
					\node[draw=#3, #4, circle, minimum size=#2*0.5*\NODESIZE, anchor=#6] at ([yshift=0.1*#2*\NODESIZE]\bdr.#6) () {};
					\node[draw=#3, #4, circle, minimum size=#2*0.5*\NODESIZE, anchor=#6] at (\bdr.#6) () {};
					\node[draw=#3, #4, circle, minimum size=#2*0.5*\NODESIZE, anchor=#6] at ([yshift=-0.1*#2*\NODESIZE]\bdr.#6) () {};
				\fi
				}
			}
		}
	},
}
\newcount\portcount

\pgfdeclareshape{fiberswitchshape}{
	\savedmacro\nin{
		\edef\nin{\pgfkeysvalueof{/tikz/fiberswitchkeys/nin}}%
	}
	\savedmacro\nout{
		\edef\nout{\pgfkeysvalueof{/tikz/fiberswitchkeys/nout}}%
	}
	\savedmacro\direction{
		\edef\direction{\pgfkeysvalueof{/tikz/fiberswitchkeys/direction}}%
	}
	\inheritsavedanchors[from=basic] 
	
	\inheritanchor[from=basic]{center}
	\inheritanchor[from=basic]{mid}		
	\inheritanchor[from=basic]{base}	
	\inheritanchor[from=basic]{north}
	\inheritanchor[from=basic]{south}		
	\inheritanchor[from=basic]{west}		
	\inheritanchor[from=basic]{mid west}				
	\inheritanchor[from=basic]{base west}		
	\inheritanchor[from=basic]{north west}		
	\inheritanchor[from=basic]{south west}		
	\inheritanchor[from=basic]{east}
	\inheritanchor[from=basic]{mid east}
	\inheritanchor[from=basic]{base east}	
	\inheritanchor[from=basic]{north east}
	\inheritanchor[from=basic]{south east}
	\inheritanchor[from=basic]{west south west}%
	\inheritanchor[from=basic]{west north west}%
	\inheritanchor[from=basic]{east north east}%
	\inheritanchor[from=basic]{east south east}%
	\inheritanchor[from=basic]{north north west}%
	\inheritanchor[from=basic]{north north east}%
	\inheritanchor[from=basic]{south south west}%
	\inheritanchor[from=basic]{south south east}%
	
	\inheritanchorborder[from=basic]	
	\inheritbackgroundpath[from=basic]

    \pgfutil@g@addto@macro\pgf@sh@s@fiberswitchshape{%
        \pgfmathsetcount{\portcount}{0}
        \pgfmathloop%
        \ifnum\the\portcount<\nin
	        \pgfutil@ifundefined{pgf@anchor@fiberswitchshape@in\the\portcount}{
		        \expandafter\xdef\csname pgf@anchor@fiberswitchshape@in\the\portcount\endcsname{%
		            \noexpand\fiberswitchshape@port[\the\portcount]{0}
		        }%
		    }{}%
	        \ifnum\the\portcount=0
    		    \pgfutil@ifundefined{pgf@anchor@fiberswitchshape@in}{%
		        \expandafter\xdef\csname pgf@anchor@fiberswitchshape@in\endcsname{%
		            \noexpand\fiberswitchshape@port[\the\portcount]{0}
		        }%
		        }{}%
		    \fi
	        \pgfmathaddtocount{\portcount}{1}	
	        \repeatpgfmathloop
        \pgfmathsetcount{\portcount}{0}
        \pgfmathloop%
    	\ifnum\the\portcount<\nout
	        \pgfutil@ifundefined{pgf@anchor@fiberswitchshape@out\the\portcount}{%
		        \expandafter\xdef\csname pgf@anchor@fiberswitchshape@out\the\portcount\endcsname{%
		            \noexpand\fiberswitchshape@port[\the\portcount]{1}
		        }%
		    }{}%
	        \ifnum\the\portcount=0
    		    \pgfutil@ifundefined{pgf@anchor@fiberswitchshape@out}{%
		        \expandafter\xdef\csname pgf@anchor@fiberswitchshape@out\endcsname{%
		            \noexpand\fiberswitchshape@port[\the\portcount]{1}
		        }%
		        }{}%
		    \fi
	        \pgfmathaddtocount{\portcount}{1}	
	        \repeatpgfmathloop
	}
}

\def\fiberswitchshape@port[#1]#2{
    \northeast \pgf@xa=\pgf@x \pgf@ya=\pgf@y
    \southwest \pgf@xb=\pgf@x \pgf@yb=\pgf@y
    
    \ifnum#2=0	
	    \if\direction\direce	
	    	\pgf@x=\pgf@xb
		    \pgf@yc=\pgf@ya \advance\pgf@yc by -\pgf@yb	
		    \pgfmathsetlength{\pgf@y}{\pgf@ya-(#1 + 0.5)*(\pgf@yc/\nin)}%
	    \fi
	    \if\direction\direcw
	    	\pgf@x=\pgf@xa
		    \pgf@yc=\pgf@ya \advance\pgf@yc by -\pgf@yb	
		    \pgfmathsetlength{\pgf@y}{\pgf@ya-(#1 + 0.5)*(\pgf@yc/\nin)}%
	    \fi
	    \if\direction\direcn
	    	\pgf@y=\pgf@yb
		    \pgf@xc=\pgf@xa \advance\pgf@xc by -\pgf@xb	
		    \pgfmathsetlength{\pgf@x}{\pgf@xb+(#1 + 0.5)*(\pgf@xc/\nin)}%
	    \fi
	    \if\direction\direcs
	    	\pgf@y=\pgf@ya
		    \pgf@xc=\pgf@xa \advance\pgf@xc by -\pgf@xb	
		    \pgfmathsetlength{\pgf@x}{\pgf@xb+(#1 + 0.5)*(\pgf@xc/\nin)}%
	    \fi
	\else	
	    \if\direction\direce	
	    	\pgf@x=\pgf@xa
		    \pgf@yc=\pgf@ya \advance\pgf@yc by -\pgf@yb	
		    \pgfmathsetlength{\pgf@y}{\pgf@ya-(#1 + 0.5)*(\pgf@yc/\nout)}%
	    \fi
	    \if\direction\direcw
	    	\pgf@x=\pgf@xb
		    \pgf@yc=\pgf@ya \advance\pgf@yc by -\pgf@yb	
		    \pgfmathsetlength{\pgf@y}{\pgf@ya-(#1 + 0.5)*(\pgf@yc/\nout)}%
	    \fi
	    \if\direction\direcn
	    	\pgf@y=\pgf@ya
		    \pgf@xc=\pgf@xa \advance\pgf@xc by -\pgf@xb	
		    \pgfmathsetlength{\pgf@x}{\pgf@xb+(#1 + 0.5)*(\pgf@xc/\nout)}%
	    \fi
	    \if\direction\direcs
	    	\pgf@y=\pgf@yb
		    \pgf@xc=\pgf@xa \advance\pgf@xc by -\pgf@xb	
		    \pgfmathsetlength{\pgf@x}{\pgf@xb+(#1 + 0.5)*(\pgf@xc/\nout)}%
	    \fi
	\fi
}

\pgfaddtoshape{fiberswitchshape}{
	\anchorlet{c}{center}		
	\anchorlet{n}{north}
	\anchorlet{e}{east}
	\anchorlet{s}{south}
	\anchorlet{w}{west}			
	\anchorlet{se}{south east}
	\anchorlet{sw}{south west}
	\anchorlet{ne}{north east}
	\anchorlet{nw}{north west}
	\anchorlet{wsw}{west south west}
	\anchorlet{wnw}{west north west}
	\anchorlet{ene}{east north east}
	\anchorlet{ese}{east south east}
	\anchorlet{nnw}{north north west}
	\anchorlet{nne}{north north east}
	\anchorlet{ssw}{south south west}
	\anchorlet{sse}{south south east}
}

\tikzset{
	/tikz/fiberswitchkeys/.cd,
	size/.initial=1,
	color/.initial=O,
	direction/.initial=e,
	linestyle/.initial={linestyle, inner sep=0.5mm},
	nin/.initial=1,	
	nout/.initial=3,
	/tikz/fiberswitch/.code={
		\pgfqkeys{/tikz/fiberswitchkeys}{#1}%
		\tikzset{/tikz/fiberswitchkeys/drawer/.expanded=%
			{\pgfkeysvalueof{/tikz/fiberswitchkeys/size}}%
			{\pgfkeysvalueof{/tikz/fiberswitchkeys/color}}%
			{\pgfkeysvalueof{/tikz/fiberswitchkeys/linestyle}}%
			{\pgfkeysvalueof{/tikz/fiberswitchkeys/nout}}%
			\if\pgfkeysvalueof{/tikz/fiberswitchkeys/direction}e
				{0}%
			\fi
			\if\pgfkeysvalueof{/tikz/fiberswitchkeys/direction}w
				{0}%
			\fi
			\if\pgfkeysvalueof{/tikz/fiberswitchkeys/direction}n
				{1}%
			\fi
			\if\pgfkeysvalueof{/tikz/fiberswitchkeys/direction}s
				{1}%
			\fi
			{\pgfkeysvalueof{/tikz/fiberswitchkeys/direction}}%
		}
	},
	/tikz/fiberswitchkeys/drawer/.code n args={6}{%
		\tikzset{
			fiberswitchshape,
			draw,
			minimum height = #1*\NODESIZE,
			minimum width = #1*\NODESIZE,
			#2,
			#3,
			append after command={
				\pgfextra{\let\bdr=\tikzlastnode%
						\ifnum#5>0
							\node[coordinate] at ($(\bdr.in)!0.25!(\bdr.out0 -| \bdr.in)$) (circlein){};
							\foreach \n [evaluate=\n as \nport using int(\n-1)] in {1,...,#4}{
								\node[coordinate] at ($(\bdr.out\nport)!0.25!(\bdr.in -| \bdr.out\nport)$) (circleout\nport){};
							}
						\else
							\node[coordinate] at ($(\bdr.in)!0.25!(\bdr.out0 |- \bdr.in)$) (circlein){};
							\foreach \n [evaluate=\n as \nport using int(\n-1)] in {1,...,#4}{
								\node[coordinate] at ($(\bdr.out\nport)!0.25!(\bdr.in |- \bdr.out\nport)$) (circleout\nport){};
							}
						\fi

						\draw[---, #2, #3, fill] (\bdr.in) to (circlein) circle (0.05);
						\foreach \n [evaluate=\n as \nport using int(\n-1)] in {1,...,#4}{
							\draw[---, #2, #3, fill] (\bdr.out\nport) to (circleout\nport) circle (0.05);
						}

						\tikzmath{
							int \nportmax;
							\nportmax = int(#4-1);
						}
						\draw[---, #2, #3] (circlein) to (circleout0){};
						\if#6e
							\draw[-->, #2, #3, looseness=0.8] ($(circlein)!0.6!(circleout0)$) to [out=-60, in=60]($(circlein)!0.6!(circleout\nportmax)$) {};
						\fi
						\if#6w
							\draw[-->, #2, #3, looseness=0.8] ($(circlein)!0.6!(circleout0)$) to [out=-120, in=120]($(circlein)!0.6!(circleout\nportmax)$) {};
						\fi
						\if#6s
							\draw[-->, #2, #3, looseness=0.8] ($(circlein)!0.6!(circleout0)$) to [out=-30, in=-150]($(circlein)!0.6!(circleout\nportmax)$) {};
						\fi
						\if#6n
							\draw[-->, #2, #3, looseness=0.8] ($(circlein)!0.6!(circleout0)$) to [out=30, in=150]($(circlein)!0.6!(circleout\nportmax)$) {};
						\fi

				}
			}
		}
	},
}
\pgfdeclareshape{filtershape}{
	\savedmacro\direction{
		\edef\direction{\pgfkeysvalueof{/tikz/filterkeys/direction}}%
	}
	\saveddimen\minwidth{
		\pgfmathsetlength\pgf@x{\pgfshapeminwidth}%
	}
	\saveddimen\minheight{
		\pgfmathsetlength\pgf@x{\pgfshapeminheight}%
	}
	\inheritsavedanchors[from=basic]

	\inheritanchor[from=basic]{center}
	\inheritanchor[from=basic]{mid}
	\inheritanchor[from=basic]{base}
	\inheritanchor[from=basic]{north}
	\inheritanchor[from=basic]{south}
	\inheritanchor[from=basic]{west}
	\inheritanchor[from=basic]{mid west}
	\inheritanchor[from=basic]{base west}
	\inheritanchor[from=basic]{north west}
	\inheritanchor[from=basic]{south west}
	\inheritanchor[from=basic]{east}
	\inheritanchor[from=basic]{mid east}
	\inheritanchor[from=basic]{base east}
	\inheritanchor[from=basic]{north east}
	\inheritanchor[from=basic]{south east}
	\inheritanchor[from=basic]{west south west}%
	\inheritanchor[from=basic]{west north west}%
	\inheritanchor[from=basic]{east north east}%
	\inheritanchor[from=basic]{east south east}%
	\inheritanchor[from=basic]{north north west}%
	\inheritanchor[from=basic]{north north east}%
	\inheritanchor[from=basic]{south south west}%
	\inheritanchor[from=basic]{south south east}%

	\inheritanchorborder[from=basic]
	\inheritbackgroundpath[from=basic]

	\pgfutil@g@addto@macro\pgf@sh@s@filtershape{%
		\pgfutil@ifundefined{pgf@anchor@filtershape@in0}{
			\expandafter\xdef\csname pgf@anchor@filtershape@in0\endcsname{%
				\noexpand\filtershape@port{0}
			}%
		}{}%
		\pgfutil@ifundefined{pgf@anchor@filtershape@in}{
			\expandafter\xdef\csname pgf@anchor@filtershape@in\endcsname{%
				\noexpand\filtershape@port{0}
			}%
		}{}%
		\pgfutil@ifundefined{pgf@anchor@filtershape@out0}{
			\expandafter\xdef\csname pgf@anchor@filtershape@out0\endcsname{%
				\noexpand\filtershape@port{1}
			}%
		}{}%
		\pgfutil@ifundefined{pgf@anchor@filtershape@out}{
			\expandafter\xdef\csname pgf@anchor@filtershape@out\endcsname{%
				\noexpand\filtershape@port{1}
			}%
		}{}%
	}
}

\def\filtershape@port#1{
	\northeast	

	\ifnum#1=0	
		\if\direction\direce
			\pgf@x=-\pgf@x
			\pgf@ya= \pgf@y
			\pgfmathsetlength{\pgf@y}{\pgf@ya-0.5*\minheight}%
		\fi
		\if\direction\direcw
			\pgf@x=\pgf@x
			\pgf@ya= \pgf@y
			\pgfmathsetlength{\pgf@y}{\pgf@ya-0.5*\minheight}%
		\fi
		\if\direction\direcn
			\pgf@y=-\pgf@y
			\pgf@xa=\pgf@x
			\pgfmathsetlength{\pgf@x}{\pgf@xa-0.5*\minwidth}%
		\fi
		\if\direction\direcs
			\pgf@y=\pgf@y
			\pgf@xa= \pgf@x
			\pgfmathsetlength{\pgf@x}{\pgf@xa-0.5*\minwidth}%
		\fi
	\else	
		\if\direction\direce
			\pgf@x=\pgf@x
			\pgf@ya= \pgf@y
			\pgfmathsetlength{\pgf@y}{\pgf@ya-0.5*\minheight}%
		\fi
		\if\direction\direcw
			\pgf@x=-\pgf@x
			\pgf@ya= \pgf@y
			\pgfmathsetlength{\pgf@y}{\pgf@ya-0.5*\minheight}%
		\fi
		\if\direction\direcn
			\pgf@y=\pgf@y
			\pgf@xa= \pgf@x
			\pgfmathsetlength{\pgf@x}{\pgf@xa-0.5*\minwidth}%
		\fi
		\if\direction\direcs
			\pgf@y=-\pgf@y
			\pgf@xa= \pgf@x
			\pgfmathsetlength{\pgf@x}{\pgf@xa-0.5*\minwidth}%
		\fi
	\fi
}

\pgfaddtoshape{filtershape}{
	\anchorlet{c}{center}
	\anchorlet{n}{north}
	\anchorlet{e}{east}
	\anchorlet{s}{south}
	\anchorlet{w}{west}
	\anchorlet{se}{south east}
	\anchorlet{sw}{south west}
	\anchorlet{ne}{north east}
	\anchorlet{nw}{north west}
	\anchorlet{wsw}{west south west}
	\anchorlet{wnw}{west north west}
	\anchorlet{ene}{east north east}
	\anchorlet{ese}{east south east}
	\anchorlet{nnw}{north north west}
	\anchorlet{nne}{north north east}
	\anchorlet{ssw}{south south west}
	\anchorlet{sse}{south south east}
}

\pgfmathsetmacro{\WSSSINEHEIGHT}{0.06}

\tikzset{
	/tikz/filterkeys/.cd,
	size/.initial=0.5,
	color/.initial=O,
	direction/.initial=e,
	linestyle/.initial={linestyle},
	fillgradient/.initial=O,
	/tikz/filter/.code={
			\pgfqkeys{/tikz/filterkeys}{#1}%
			\tikzset{/tikz/filterkeys/drawer/.expanded=%
					{\pgfkeysvalueof{/tikz/filterkeys/direction}}%
					{\pgfkeysvalueof{/tikz/filterkeys/size}}%
					{\pgfkeysvalueof{/tikz/filterkeys/color}}%
					{\pgfkeysvalueof{/tikz/filterkeys/linestyle}}%
					{\pgfkeysvalueof{/tikz/filterkeys/fillgradient}}%
			}
		},
	/tikz/filterkeys/drawer/.code n args={5}{%
			\tikzset{
				filtershape,
				minimum height=#2*\NODESIZE,
				minimum width=#2*\NODESIZE,
				#3,
				#4,
				draw,
				append after command={
						\pgfextra{\let\bdr=\tikzlastnode%
							\node[#5, fit=(\bdr.nw)(\bdr.se)] (boxgradient){};

							\node[coordinate] at (\bdr.wnw -| \bdr.nnw) (hl){};
							\node[coordinate] at (\bdr.ene -| \bdr.nne) (hr){};
							\draw[#3,---, rounded corners = 0] (hl) sin ($(hl)!0.25!(hr) + (0,0.002*#2*\NODESIZE)$) cos ($(hl)!0.5!(hr)$) sin ($(hl)!0.75!(hr) + (0,-0.002*#2*\NODESIZE)$) cos (hr);

							\node[coordinate] at (\bdr.w -| \bdr.nnw) (ml){};
							\node[coordinate] at (\bdr.e -| \bdr.nne) (mr){};
							\draw[#3,---, rounded corners = 0] (ml) sin ($(ml)!0.25!(mr) + (0,0.002*#2*\NODESIZE)$) cos ($(ml)!0.5!(mr)$) sin ($(ml)!0.75!(mr) + (0,-0.002*#2*\NODESIZE)$) cos (mr);

							\node[coordinate] at (\bdr.wsw -| \bdr.nnw) (ll){};
							\node[coordinate] at (\bdr.ese -| \bdr.nne) (lr){};
							\draw[#3,---, rounded corners = 0] (ll) sin ($(ll)!0.25!(lr) + (0,0.002*#2*\NODESIZE)$) cos ($(ll)!0.5!(lr)$) sin ($(ll)!0.75!(lr) + (0,-0.002*#2*\NODESIZE)$) cos (lr);

							\draw[#3, ---] ($(hl)!0.25!(hr) - (0,0.002*#2*\NODESIZE)$) -- ($(hl)!0.75!(hr) + (0,0.002*#2*\NODESIZE)$);
							\draw[#3, ---] ($(ll)!0.25!(lr) - (0,0.002*#2*\NODESIZE)$) -- ($(ll)!0.75!(lr) + (0,0.002*#2*\NODESIZE)$);
						}
					}
			}
		},
}
\newcount\portcount

\pgfdeclareshape{lensshape}{
	\savedmacro\nport{
		\edef\nport{\pgfkeysvalueof{/tikz/lenskeys/nport}}%
	}
	\inheritsavedanchors[from=basic] 
	
	\inheritanchor[from=basic]{center}
	\inheritanchor[from=basic]{mid}		
	\inheritanchor[from=basic]{base}	
	\inheritanchor[from=basic]{north}
	\inheritanchor[from=basic]{south}		
	\inheritanchor[from=basic]{west}		
	\inheritanchor[from=basic]{mid west}				
	\inheritanchor[from=basic]{base west}		
	\inheritanchor[from=basic]{north west}		
	\inheritanchor[from=basic]{south west}		
	\inheritanchor[from=basic]{east}
	\inheritanchor[from=basic]{mid east}
	\inheritanchor[from=basic]{base east}	
	\inheritanchor[from=basic]{north east}
	\inheritanchor[from=basic]{south east}
	\inheritanchor[from=basic]{west south west}%
	\inheritanchor[from=basic]{west north west}%
	\inheritanchor[from=basic]{east north east}%
	\inheritanchor[from=basic]{east south east}%
	\inheritanchor[from=basic]{north north west}%
	\inheritanchor[from=basic]{north north east}%
	\inheritanchor[from=basic]{south south west}%
	\inheritanchor[from=basic]{south south east}%
	
	\inheritanchorborder[from=basic]	
	\inheritbackgroundpath[from=basic]

    \pgfutil@g@addto@macro\pgf@sh@s@lensshape{%
        \pgfmathsetcount{\portcount}{0}
        \pgfmathloop%
        \ifnum\the\portcount<\nport
	        \pgfutil@ifundefined{pgf@anchor@lensshape@in\the\portcount}{
		        \expandafter\xdef\csname pgf@anchor@lensshape@in\the\portcount\endcsname{%
		            \noexpand\lensshape@port[\the\portcount]
		        }%
		    }{}%
	        \ifnum\the\portcount=0
    		    \pgfutil@ifundefined{pgf@anchor@lensshape@in}{%
		        \expandafter\xdef\csname pgf@anchor@lensshape@in\endcsname{%
		            \noexpand\lensshape@port[\the\portcount]
		        }%
		        }{}%
		    \fi
		    \pgfutil@ifundefined{pgf@anchor@lensshape@out\the\portcount}{%
		        \expandafter\xdef\csname pgf@anchor@lensshape@out\the\portcount\endcsname{%
		            \noexpand\lensshape@port[\the\portcount]
		        }%
		    }{}%
	        \ifnum\the\portcount=0
    		    \pgfutil@ifundefined{pgf@anchor@lensshape@out}{%
		        \expandafter\xdef\csname pgf@anchor@lensshape@out\endcsname{%
		            \noexpand\lensshape@port[\the\portcount]
		        }%
		        }{}%
		    \fi
	        \pgfmathaddtocount{\portcount}{1}	
	        \repeatpgfmathloop
	}
}

\def\lensshape@port[#1]{
    \northeast \pgf@xa=\pgf@x \pgf@ya=\pgf@y
    \southwest \pgf@xb=\pgf@x \pgf@yb=\pgf@y
    
	\pgfmathsetlength{\pgf@x}{0.5*\pgf@xa + 0.5*\pgf@xb}%
    \pgf@yc=\pgf@ya \advance\pgf@yc by -\pgf@yb	
    \pgfmathsetlength{\pgf@y}{\pgf@ya-(#1 + 0.5)*(\pgf@yc/\nport)}%
}

\pgfaddtoshape{lensshape}{
	\anchorlet{c}{center}		
	\anchorlet{n}{north}
	\anchorlet{e}{east}
	\anchorlet{s}{south}
	\anchorlet{w}{west}			
	\anchorlet{se}{south east}
	\anchorlet{sw}{south west}
	\anchorlet{ne}{north east}
	\anchorlet{nw}{north west}
	\anchorlet{wsw}{west south west}
	\anchorlet{wnw}{west north west}
	\anchorlet{ene}{east north east}
	\anchorlet{ese}{east south east}
	\anchorlet{nnw}{north north west}
	\anchorlet{nne}{north north east}
	\anchorlet{ssw}{south south west}
	\anchorlet{sse}{south south east}
}

\tikzset{
	/tikz/lenskeys/.cd,
	height/.initial=1,
	width/.initial=0.3,	
	color/.initial=O,
	rotation/.initial=0,
	linestyle/.initial={linestyle, inner sep=0.5mm},
	nport/.initial=1,
	/tikz/lens/.code={
		\pgfqkeys{/tikz/lenskeys}{#1}%
		\tikzset{/tikz/lenskeys/drawer/.expanded=%
			{\pgfkeysvalueof{/tikz/lenskeys/height}}%
			{\pgfkeysvalueof{/tikz/lenskeys/width}}%
			{\pgfkeysvalueof{/tikz/lenskeys/color}}%
			{\pgfkeysvalueof{/tikz/lenskeys/linestyle}}%
			{\pgfkeysvalueof{/tikz/lenskeys/rotation}}%
		}
	},
	/tikz/lenskeys/drawer/.code n args={5}{%
		\tikzset{
			lensshape,
			rotate=#5,
			minimum height=#1*\NODESIZE,
			minimum width=#2*\NODESIZE,
			append after command={
				\pgfextra{\let\bdr=\tikzlastnode%
					\draw[---, #3, #4, rounded corners = 0] (\bdr.n) to [in=120+#5, out=240+#5] (\bdr.s) to [in=-60+#5, out=60+#5] (\bdr.n) -- cycle;
				}
			}
		}
	},
}

\newcount\portcount

\pgfdeclareshape{mirrorshape}{
	\savedmacro\nport{
		\edef\nport{\pgfkeysvalueof{/tikz/mirrorkeys/nport}}%
	}
	\inheritsavedanchors[from=basic] 
	
	\inheritanchor[from=basic]{center}
	\inheritanchor[from=basic]{mid}		
	\inheritanchor[from=basic]{base}	
	\inheritanchor[from=basic]{north}
	\inheritanchor[from=basic]{south}		
	\inheritanchor[from=basic]{west}		
	\inheritanchor[from=basic]{mid west}				
	\inheritanchor[from=basic]{base west}		
	\inheritanchor[from=basic]{north west}		
	\inheritanchor[from=basic]{south west}		
	\inheritanchor[from=basic]{east}
	\inheritanchor[from=basic]{mid east}
	\inheritanchor[from=basic]{base east}	
	\inheritanchor[from=basic]{north east}
	\inheritanchor[from=basic]{south east}
	\inheritanchor[from=basic]{west south west}%
	\inheritanchor[from=basic]{west north west}%
	\inheritanchor[from=basic]{east north east}%
	\inheritanchor[from=basic]{east south east}%
	\inheritanchor[from=basic]{north north west}%
	\inheritanchor[from=basic]{north north east}%
	\inheritanchor[from=basic]{south south west}%
	\inheritanchor[from=basic]{south south east}%
	
	\inheritanchorborder[from=basic]	
	\inheritbackgroundpath[from=basic]

    \pgfutil@g@addto@macro\pgf@sh@s@mirrorshape{%
        \pgfmathsetcount{\portcount}{0}
        \pgfmathloop%
        \ifnum\the\portcount<\nport
	        \pgfutil@ifundefined{pgf@anchor@mirrorshape@in\the\portcount}{
		        \expandafter\xdef\csname pgf@anchor@mirrorshape@in\the\portcount\endcsname{%
		            \noexpand\mirrorshape@port[\the\portcount]
		        }%
		    }{}%
	        \ifnum\the\portcount=0
    		    \pgfutil@ifundefined{pgf@anchor@mirrorshape@in}{%
		        \expandafter\xdef\csname pgf@anchor@mirrorshape@in\endcsname{%
		            \noexpand\mirrorshape@port[\the\portcount]
		        }%
		        }{}%
		    \fi
		    \pgfutil@ifundefined{pgf@anchor@mirrorshape@out\the\portcount}{%
		        \expandafter\xdef\csname pgf@anchor@mirrorshape@out\the\portcount\endcsname{%
		            \noexpand\mirrorshape@port[\the\portcount]
		        }%
		    }{}%
	        \ifnum\the\portcount=0
    		    \pgfutil@ifundefined{pgf@anchor@mirrorshape@out}{%
		        \expandafter\xdef\csname pgf@anchor@mirrorshape@out\endcsname{%
		            \noexpand\mirrorshape@port[\the\portcount]
		        }%
		        }{}%
		    \fi
	        \pgfmathaddtocount{\portcount}{1}	
	        \repeatpgfmathloop
	}
}

\def\mirrorshape@port[#1]{
    \northeast \pgf@xa=\pgf@x \pgf@ya=\pgf@y
    \southwest \pgf@xb=\pgf@x \pgf@yb=\pgf@y
    
	\pgf@x=\pgf@xb
    \pgf@yc=\pgf@ya \advance\pgf@yc by -\pgf@yb	
    \pgfmathsetlength{\pgf@y}{\pgf@ya-(#1 + 0.5)*(\pgf@yc/\nport)}%
}

\pgfaddtoshape{mirrorshape}{
	\anchorlet{c}{center}		
	\anchorlet{n}{north}
	\anchorlet{e}{east}
	\anchorlet{s}{south}
	\anchorlet{w}{west}			
	\anchorlet{se}{south east}
	\anchorlet{sw}{south west}
	\anchorlet{ne}{north east}
	\anchorlet{nw}{north west}
	\anchorlet{wsw}{west south west}
	\anchorlet{wnw}{west north west}
	\anchorlet{ene}{east north east}
	\anchorlet{ese}{east south east}
	\anchorlet{nnw}{north north west}
	\anchorlet{nne}{north north east}
	\anchorlet{ssw}{south south west}
	\anchorlet{sse}{south south east}
}

\tikzset{
	/tikz/mirrorkeys/.cd,
	height/.initial=1,
	width/.initial=0.15,	
	color/.initial=O,
	rotation/.initial=0,
	linestyle/.initial={linestyle, inner sep=0.5mm},
	nport/.initial=1,
	nlines/.initial=5,
	/tikz/mirror/.code={
		\pgfqkeys{/tikz/mirrorkeys}{#1}%
		\tikzset{/tikz/mirrorkeys/drawer/.expanded=%
			{\pgfkeysvalueof{/tikz/mirrorkeys/height}}%
			{\pgfkeysvalueof{/tikz/mirrorkeys/width}}%
			{\pgfkeysvalueof{/tikz/mirrorkeys/color}}%
			{\pgfkeysvalueof{/tikz/mirrorkeys/linestyle}}%
			{\pgfkeysvalueof{/tikz/mirrorkeys/rotation}}%
			{\pgfkeysvalueof{/tikz/mirrorkeys/nlines}}%
		}
	},
	/tikz/mirrorkeys/drawer/.code n args={6}{%
		\tikzset{
			mirrorshape,
			rotate=#5,
			minimum height=#1*\NODESIZE,
			minimum width=#2*\NODESIZE,
			append after command={
				\pgfextra{\let\bdr=\tikzlastnode%
					\draw[---, #3, #4] (\bdr.nw) to (\bdr.sw){};
					\foreach \nline [evaluate=\nline as \linepos using (\nline-0.8)/(#6-0.6)] in {1,...,#6}{
						\draw[---, #3, #4] ($(\bdr.nw)!\linepos!(\bdr.sw)$) to +(#5-45:#2*1.41421356237*\NODESIZE pt){};
					}
				}
			}
		}
	},
}
\pgfdeclareshape{muxshape}{
	\savedmacro\nports{
		\edef\nports{\pgfkeysvalueof{/tikz/muxkeys/nports}}%
	}
	\savedmacro\direction{
		\edef\direction{\pgfkeysvalueof{/tikz/muxkeys/direction}}%
	}
	\savedmacro\inverted{
		\edef\inverted{\pgfkeysvalueof{/tikz/muxkeys/inverted}}%
	}
	\savedmacro\ninports{
		\ifnum\inverted=0
			\edef\ninports{\nports}
		\else
			\edef\ninports{1}%
		\fi
	}
	\savedmacro\noutports{
		\ifnum\inverted=0
			\edef\noutports{1}%
		\else
			\edef\noutports{\nports}%
		\fi
	}
	\inheritsavedanchors[from=basic]

	\inheritanchor[from=basic]{center}
	\inheritanchor[from=basic]{mid}
	\inheritanchor[from=basic]{base}
	\inheritanchor[from=basic]{north}
	\inheritanchor[from=basic]{south}
	\inheritanchor[from=basic]{west}
	\inheritanchor[from=basic]{mid west}
	\inheritanchor[from=basic]{base west}
	\inheritanchor[from=basic]{north west}
	\inheritanchor[from=basic]{south west}
	\inheritanchor[from=basic]{east}
	\inheritanchor[from=basic]{mid east}
	\inheritanchor[from=basic]{base east}
	\inheritanchor[from=basic]{north east}
	\inheritanchor[from=basic]{south east}
	\inheritanchor[from=basic]{west south west}%
	\inheritanchor[from=basic]{west north west}%
	\inheritanchor[from=basic]{east north east}%
	\inheritanchor[from=basic]{east south east}%
	\inheritanchor[from=basic]{north north west}%
	\inheritanchor[from=basic]{north north east}%
	\inheritanchor[from=basic]{south south west}%
	\inheritanchor[from=basic]{south south east}%

	\inheritanchorborder[from=basic]
	\inheritbackgroundpath[from=basic]

	\pgfutil@g@addto@macro\pgf@sh@s@muxshape{%
		\pgfmathsetcount{\portcount}{0}
		\pgfmathloop%
		\ifnum\the\portcount<\nports
		\ifnum\the\portcount<\ninports
			\pgfutil@ifundefined{pgf@anchor@muxshape@in\the\portcount}{
				\expandafter\xdef\csname pgf@anchor@muxshape@in\the\portcount\endcsname{%
					\noexpand\muxshape@port[\the\portcount]{0}
				}%
			}{}%
			\ifnum\the\portcount=0
				\pgfutil@ifundefined{pgf@anchor@muxshape@in}{%
					\expandafter\xdef\csname pgf@anchor@muxshape@in\endcsname{%
						\noexpand\muxshape@port[\the\portcount]{0}
					}%
				}{}%
			\fi
		\fi
		\ifnum\the\portcount<\noutports
			\pgfutil@ifundefined{pgf@anchor@muxshape@out\the\portcount}{%
				\expandafter\xdef\csname pgf@anchor@muxshape@out\the\portcount\endcsname{%
					\noexpand\muxshape@port[\the\portcount]{1}
				}%
			}{}%
			\ifnum\the\portcount=0
				\pgfutil@ifundefined{pgf@anchor@muxshape@out}{%
					\expandafter\xdef\csname pgf@anchor@muxshape@out\endcsname{%
						\noexpand\muxshape@port[\the\portcount]{1}
					}%
				}{}%
			\fi
		\fi
		\pgfmathaddtocount{\portcount}{1}	
		\repeatpgfmathloop
	}
}

\def\muxshape@port[#1]#2{
	\northeast \pgf@xa=\pgf@x \pgf@ya=\pgf@y
	\southwest \pgf@xb=\pgf@x \pgf@yb=\pgf@y

	\ifnum#2=0
		\ifnum\inverted=0
			\def\chooseports{0}
		\else
			\def\chooseports{1}
		\fi
	\else
		\ifnum\inverted=0
			\def\chooseports{1}
		\else
			\def\chooseports{0}
		\fi
	\fi

	\ifnum\chooseports=0	
		\if\direction\direce
			\pgf@x=\pgf@xb
			\pgf@yc=\pgf@ya \advance\pgf@yc by -\pgf@yb	
			\pgfmathsetlength{\pgf@y}{\pgf@ya-(#1 + 0.5)*(\pgf@yc/\nports)}%
		\fi
		\if\direction\direcw
			\pgf@x=\pgf@xa
			\pgf@yc=\pgf@ya \advance\pgf@yc by -\pgf@yb	
			\pgfmathsetlength{\pgf@y}{\pgf@ya-(#1 + 0.5)*(\pgf@yc/\nports)}%
		\fi
		\if\direction\direcn
			\pgf@y=\pgf@yb
			\pgf@xc=\pgf@xa \advance\pgf@xc by -\pgf@xb	
			\pgfmathsetlength{\pgf@x}{\pgf@xb+(#1 + 0.5)*(\pgf@xc/\nports)}%
		\fi
		\if\direction\direcs
			\pgf@y=\pgf@ya
			\pgf@xc=\pgf@xa \advance\pgf@xc by -\pgf@xb	
			\pgfmathsetlength{\pgf@x}{\pgf@xb+(#1 + 0.5)*(\pgf@xc/\nports)}%
		\fi
	\else	
		\if\direction\direce
			\pgf@x=\pgf@xa
			\pgf@yc=\pgf@ya \advance\pgf@yc by -\pgf@yb	
			\pgfmathsetlength{\pgf@y}{\pgf@ya-0.5\pgf@yc}%
		\fi
		\if\direction\direcw
			\pgf@x=\pgf@xb
			\pgf@yc=\pgf@ya \advance\pgf@yc by -\pgf@yb	
			\pgfmathsetlength{\pgf@y}{\pgf@ya-0.5\pgf@yc}%
		\fi
		\if\direction\direcn
			\pgf@y=\pgf@ya
			\pgf@xc=\pgf@xa \advance\pgf@xc by -\pgf@xb	
			\pgfmathsetlength{\pgf@x}{\pgf@xa-0.5\pgf@xc}%
		\fi
		\if\direction\direcs
			\pgf@y=\pgf@yb
			\pgf@xc=\pgf@xa \advance\pgf@xc by -\pgf@xb	
			\pgfmathsetlength{\pgf@x}{\pgf@xa-0.5\pgf@xc}%
		\fi
	\fi
}

\pgfaddtoshape{muxshape}{
	\anchorlet{c}{center}
	\anchorlet{n}{north}
	\anchorlet{e}{east}
	\anchorlet{s}{south}
	\anchorlet{w}{west}
	\anchorlet{se}{south east}
	\anchorlet{sw}{south west}
	\anchorlet{ne}{north east}
	\anchorlet{nw}{north west}
	\anchorlet{wsw}{west south west}
	\anchorlet{wnw}{west north west}
	\anchorlet{ene}{east north east}
	\anchorlet{ese}{east south east}
	\anchorlet{nnw}{north north west}
	\anchorlet{nne}{north north east}
	\anchorlet{ssw}{south south west}
	\anchorlet{sse}{south south east}
}

\tikzset{
/tikz/muxkeys/.cd,
height/.initial=1,
width/.initial=0.5,
color/.initial=O,
direction/.initial=e,
linestyle/.initial={linestyle, rounded corners = 0},
nports/.initial=3,
inverted/.initial=0,	
smux/.initial=0,
angle/.initial=60,
fillgradient/.initial=O,
/tikz/mux/.code={
\pgfqkeys{/tikz/muxkeys}{#1}%
\tikzset{/tikz/muxkeys/drawer/.expanded=%
\if\pgfkeysvalueof{/tikz/muxkeys/direction}e
	{\pgfkeysvalueof{/tikz/muxkeys/width}}%
	{\pgfkeysvalueof{/tikz/muxkeys/height}}%
	{0}%
	{-90}%
\fi
\if\pgfkeysvalueof{/tikz/muxkeys/direction}w
	{\pgfkeysvalueof{/tikz/muxkeys/width}}%
	{\pgfkeysvalueof{/tikz/muxkeys/height}}%
	{0}%
	{90}%
\fi
\if\pgfkeysvalueof{/tikz/muxkeys/direction}n
	{\pgfkeysvalueof{/tikz/muxkeys/width}}%
	{\pgfkeysvalueof{/tikz/muxkeys/height}}%
	{1}%
	{0}%
\fi
\if\pgfkeysvalueof{/tikz/muxkeys/direction}s
	{\pgfkeysvalueof{/tikz/muxkeys/width}}%
	{\pgfkeysvalueof{/tikz/muxkeys/height}}%
	{1}%
	{180}%
\fi
{\pgfkeysvalueof{/tikz/muxkeys/color}}%
{\pgfkeysvalueof{/tikz/muxkeys/linestyle}}%
{\pgfkeysvalueof{/tikz/muxkeys/smux}}%
{\pgfkeysvalueof{/tikz/muxkeys/angle}}%
{\pgfkeysvalueof{/tikz/muxkeys/fillgradient}}%
}
},
/tikz/muxkeys/drawer/.code n args={9}{%
		\tikzset{
			muxshape,
			#6,
			#5,
			minimum height=
			\ifnum#3>0	
				#1*\NODESIZE
			\else
				#2*\NODESIZE
			\fi
			,minimum width=
			\ifnum#3>0
				#2*\NODESIZE
			\else
				#1*\NODESIZE
			\fi
			,append after command={
					\pgfextra{\let\bdr=\tikzlastnode%
						\node[trapezium, line width = \NODETHICKNESS, minimum height=#1*\NODESIZE, minimum width=#2*\NODESIZE, trapezium stretches=true, rotate=#4, trapezium angle=#8, inner sep=0.001mm] at (\bdr) (trap) {};	
						\ifnum#7=0
							\draw[#9, #5, #6] (trap.bottom left corner) to (trap.top left corner) to (trap.top right corner) to (trap.bottom right corner) to cycle;
						\else
							\draw[#9, #5, #6] (trap.bottom left corner) to[in=#4-90, out=#4+90] (trap.top left corner) to (trap.top right corner) to[in=#4+90,out=#4-90] (trap.bottom right corner) to cycle;
						\fi
					}
				}
		}
	},
}
\newcount\portcount

\pgfdeclareshape{polswitchshape}{
	\savedmacro\nin{
		\edef\nin{\pgfkeysvalueof{/tikz/polswitchkeys/nin}}%
	}
	\savedmacro\nout{
		\edef\nout{\pgfkeysvalueof{/tikz/polswitchkeys/nout}}%
	}
	\savedmacro\direction{
		\edef\direction{\pgfkeysvalueof{/tikz/polswitchkeys/direction}}%
	}
	\inheritsavedanchors[from=basic] 
	
	\inheritanchor[from=basic]{center}
	\inheritanchor[from=basic]{mid}		
	\inheritanchor[from=basic]{base}	
	\inheritanchor[from=basic]{north}
	\inheritanchor[from=basic]{south}		
	\inheritanchor[from=basic]{west}		
	\inheritanchor[from=basic]{mid west}				
	\inheritanchor[from=basic]{base west}		
	\inheritanchor[from=basic]{north west}		
	\inheritanchor[from=basic]{south west}		
	\inheritanchor[from=basic]{east}
	\inheritanchor[from=basic]{mid east}
	\inheritanchor[from=basic]{base east}	
	\inheritanchor[from=basic]{north east}
	\inheritanchor[from=basic]{south east}
	\inheritanchor[from=basic]{west south west}%
	\inheritanchor[from=basic]{west north west}%
	\inheritanchor[from=basic]{east north east}%
	\inheritanchor[from=basic]{east south east}%
	\inheritanchor[from=basic]{north north west}%
	\inheritanchor[from=basic]{north north east}%
	\inheritanchor[from=basic]{south south west}%
	\inheritanchor[from=basic]{south south east}%
	
	\inheritanchorborder[from=basic]	
	\inheritbackgroundpath[from=basic]

    \pgfutil@g@addto@macro\pgf@sh@s@polswitchshape{%
        \pgfmathsetcount{\portcount}{0}
        \pgfmathloop%
        \ifnum\the\portcount<\nin
	        \pgfutil@ifundefined{pgf@anchor@polswitchshape@in\the\portcount}{
		        \expandafter\xdef\csname pgf@anchor@polswitchshape@in\the\portcount\endcsname{%
		            \noexpand\polswitchshape@port[\the\portcount]{0}
		        }%
		    }{}%
	        \ifnum\the\portcount=0
    		    \pgfutil@ifundefined{pgf@anchor@polswitchshape@in}{%
		        \expandafter\xdef\csname pgf@anchor@polswitchshape@in\endcsname{%
		            \noexpand\polswitchshape@port[\the\portcount]{0}
		        }%
		        }{}%
		    \fi
	        \pgfmathaddtocount{\portcount}{1}	
	        \repeatpgfmathloop
        \pgfmathsetcount{\portcount}{0}
        \pgfmathloop%
    	\ifnum\the\portcount<\nout
	        \pgfutil@ifundefined{pgf@anchor@polswitchshape@out\the\portcount}{%
		        \expandafter\xdef\csname pgf@anchor@polswitchshape@out\the\portcount\endcsname{%
		            \noexpand\polswitchshape@port[\the\portcount]{1}
		        }%
		    }{}%
	        \ifnum\the\portcount=0
    		    \pgfutil@ifundefined{pgf@anchor@polswitchshape@out}{%
		        \expandafter\xdef\csname pgf@anchor@polswitchshape@out\endcsname{%
		            \noexpand\polswitchshape@port[\the\portcount]{1}
		        }%
		        }{}%
		    \fi
	        \pgfmathaddtocount{\portcount}{1}	
	        \repeatpgfmathloop
	}
}

\def\polswitchshape@port[#1]#2{
    \northeast \pgf@xa=\pgf@x \pgf@ya=\pgf@y
    \southwest \pgf@xb=\pgf@x \pgf@yb=\pgf@y
    
    \ifnum#2=0	
	    \if\direction\direce	
	    	\pgf@x=\pgf@xb
		    \pgf@yc=\pgf@ya \advance\pgf@yc by -\pgf@yb	
		    \pgfmathsetlength{\pgf@y}{\pgf@ya-(#1 + 0.5)*(\pgf@yc/\nin)}%
	    \fi
	    \if\direction\direcw
	    	\pgf@x=\pgf@xa
		    \pgf@yc=\pgf@ya \advance\pgf@yc by -\pgf@yb	
		    \pgfmathsetlength{\pgf@y}{\pgf@ya-(#1 + 0.5)*(\pgf@yc/\nin)}%
	    \fi
	    \if\direction\direcn
	    	\pgf@y=\pgf@yb
		    \pgf@xc=\pgf@xa \advance\pgf@xc by -\pgf@xb	
		    \pgfmathsetlength{\pgf@x}{\pgf@xb+(#1 + 0.5)*(\pgf@xc/\nin)}%
	    \fi
	    \if\direction\direcs
	    	\pgf@y=\pgf@ya
		    \pgf@xc=\pgf@xa \advance\pgf@xc by -\pgf@xb	
		    \pgfmathsetlength{\pgf@x}{\pgf@xb+(#1 + 0.5)*(\pgf@xc/\nin)}%
	    \fi
	\else	
	    \if\direction\direce	
	    	\pgf@x=\pgf@xa
		    \pgf@yc=\pgf@ya \advance\pgf@yc by -\pgf@yb	
		    \pgfmathsetlength{\pgf@y}{\pgf@ya-(#1 + 0.5)*(\pgf@yc/\nout)}%
	    \fi
	    \if\direction\direcw
	    	\pgf@x=\pgf@xb
		    \pgf@yc=\pgf@ya \advance\pgf@yc by -\pgf@yb	
		    \pgfmathsetlength{\pgf@y}{\pgf@ya-(#1 + 0.5)*(\pgf@yc/\nout)}%
	    \fi
	    \if\direction\direcn
	    	\pgf@y=\pgf@ya
		    \pgf@xc=\pgf@xa \advance\pgf@xc by -\pgf@xb	
		    \pgfmathsetlength{\pgf@x}{\pgf@xb+(#1 + 0.5)*(\pgf@xc/\nout)}%
	    \fi
	    \if\direction\direcs
	    	\pgf@y=\pgf@yb
		    \pgf@xc=\pgf@xa \advance\pgf@xc by -\pgf@xb	
		    \pgfmathsetlength{\pgf@x}{\pgf@xb+(#1 + 0.5)*(\pgf@xc/\nout)}%
	    \fi
	\fi
}

\pgfaddtoshape{polswitchshape}{
	\anchorlet{c}{center}		
	\anchorlet{n}{north}
	\anchorlet{e}{east}
	\anchorlet{s}{south}
	\anchorlet{w}{west}			
	\anchorlet{se}{south east}
	\anchorlet{sw}{south west}
	\anchorlet{ne}{north east}
	\anchorlet{nw}{north west}
	\anchorlet{wsw}{west south west}
	\anchorlet{wnw}{west north west}
	\anchorlet{ene}{east north east}
	\anchorlet{ese}{east south east}
	\anchorlet{nnw}{north north west}
	\anchorlet{nne}{north north east}
	\anchorlet{ssw}{south south west}
	\anchorlet{sse}{south south east}
}

\tikzset{
	/tikz/polswitchkeys/.cd,
	size/.initial=1,
	color/.initial=O,
	direction/.initial=e,
	linestyle/.initial={linestyle, inner sep=0.5mm},
	nin/.initial=1,	
	nout/.initial=1, 
	/tikz/polswitch/.code={
		\pgfqkeys{/tikz/polswitchkeys}{#1}%
		\tikzset{/tikz/polswitchkeys/drawer/.expanded=%
			{\pgfkeysvalueof{/tikz/polswitchkeys/size}}%
			{\pgfkeysvalueof{/tikz/polswitchkeys/color}}%
			{\pgfkeysvalueof{/tikz/polswitchkeys/linestyle}}%
			{\pgfkeysvalueof{/tikz/polswitchkeys/nout}}%
			\if\pgfkeysvalueof{/tikz/polswitchkeys/direction}e
				{0}%
			\fi
			\if\pgfkeysvalueof{/tikz/polswitchkeys/direction}w
				{0}%
			\fi
			\if\pgfkeysvalueof{/tikz/polswitchkeys/direction}n
				{1}%
			\fi
			\if\pgfkeysvalueof{/tikz/polswitchkeys/direction}s
				{1}%
			\fi
			{\pgfkeysvalueof{/tikz/polswitchkeys/direction}}%
		}
	},
	/tikz/polswitchkeys/drawer/.code n args={6}{%
		\tikzset{
			polswitchshape,
			draw,
			minimum height = #1*\NODESIZE,
			minimum width = #1*\NODESIZE,
			#2,
			#3,
			append after command={
				\pgfextra{\let\bdr=\tikzlastnode%
						\node[coordinate] at ($(\bdr.in)!0.25!(\bdr.out)$) (circlein){};
						\node[coordinate] at ($(\bdr.in)!0.75!(\bdr.out)$) (circleout){};

						\draw[---, #2, #3, fill] (\bdr.in) to (circlein) circle (0.05);
						\draw[---, #2, #3, fill] (\bdr.out) to (circleout) circle (0.05);

						\node[coordinate] at ($(\bdr.in)!0.6!(\bdr.out)$) (circlemiddle){};
						\ifnum#5>0
							\node[coordinate] at ($(circlemiddle)!0.5!(circlemiddle -| \bdr.e)$) (circletopcor){};
							\node[coordinate] at ($(circlemiddle)!0.5!(circlemiddle -| \bdr.w)$) (circlebotcor){};
						\else
							\node[coordinate] at ($(circlemiddle)!0.5!(circlemiddle |- \bdr.n)$) (circletopcor){};
							\node[coordinate] at ($(circlemiddle)!0.5!(circlemiddle |- \bdr.s)$) (circlebotcor){};
						\fi

						\node[draw, circle, #2, #3, minimum size=0.3*\FNODESIZE] at (circletopcor) (circletop){};
						\node[draw, circle, #2, #3, minimum size=0.3*\FNODESIZE] at (circlebotcor) (circlebot){};
						\draw[-->, #2, #3] (circletop.south) to (circletop.north){};
						\draw[-->, #2, #3] (circlebot.west) to (circlebot.east){};

						\if#6e
							\draw[-->, #2, #3] ([xshift=-0.05*\FNODESIZE]circletop.south west) to [out=-120, in=120] ([xshift=-0.05*\FNODESIZE]circlebot.north west){};
						\fi
						\if#6w
							\draw[-->, #2, #3] ([xshift=0.05*\FNODESIZE]circletop.south east) to [out=-60, in=60] ([xshift=0.05*\FNODESIZE]circlebot.north east){};
						\fi
						\if#6s
							\draw[-->, #2, #3] ([yshift=0.05*\FNODESIZE]circletop.north west) to [out=150, in=30] ([yshift=0.05*\FNODESIZE]circlebot.north east){};
						\fi
						\if#6n
							\draw[-->, #2, #3] ([yshift=-0.05*\FNODESIZE]circletop.south west) to [out=-150, in=-30] ([yshift=-0.05*\FNODESIZE]circlebot.south east){};
						\fi

				}
			}
		}
	},
}
\pgfdeclareshape{pdshape}{
	\savedmacro\direction{
		\edef\direction{\pgfkeysvalueof{/tikz/pdkeys/direction}}%
	}
	\saveddimen\minwidth{
		\pgfmathsetlength\pgf@x{\pgfshapeminwidth}%
	}
	\saveddimen\minheight{
		\pgfmathsetlength\pgf@x{\pgfshapeminheight}%
	}
	\inheritsavedanchors[from=basic]

	\inheritanchor[from=basic]{center}
	\inheritanchor[from=basic]{mid}
	\inheritanchor[from=basic]{base}
	\inheritanchor[from=basic]{north}
	\inheritanchor[from=basic]{south}
	\inheritanchor[from=basic]{west}
	\inheritanchor[from=basic]{mid west}
	\inheritanchor[from=basic]{base west}
	\inheritanchor[from=basic]{north west}
	\inheritanchor[from=basic]{south west}
	\inheritanchor[from=basic]{east}
	\inheritanchor[from=basic]{mid east}
	\inheritanchor[from=basic]{base east}
	\inheritanchor[from=basic]{north east}
	\inheritanchor[from=basic]{south east}
	\inheritanchor[from=basic]{west south west}%
	\inheritanchor[from=basic]{west north west}%
	\inheritanchor[from=basic]{east north east}%
	\inheritanchor[from=basic]{east south east}%
	\inheritanchor[from=basic]{north north west}%
	\inheritanchor[from=basic]{north north east}%
	\inheritanchor[from=basic]{south south west}%
	\inheritanchor[from=basic]{south south east}%

	\inheritanchorborder[from=basic]
	\inheritbackgroundpath[from=basic]

	\pgfutil@g@addto@macro\pgf@sh@s@pdshape{%
		\pgfutil@ifundefined{pgf@anchor@pdshape@in0}{
			\expandafter\xdef\csname pgf@anchor@pdshape@in0\endcsname{%
				\noexpand\pdshape@port{0}
			}%
		}{}%
		\pgfutil@ifundefined{pgf@anchor@pdshape@in}{
			\expandafter\xdef\csname pgf@anchor@pdshape@in\endcsname{%
				\noexpand\pdshape@port{0}
			}%
		}{}%
		\pgfutil@ifundefined{pgf@anchor@pdshape@out0}{
			\expandafter\xdef\csname pgf@anchor@pdshape@out0\endcsname{%
				\noexpand\pdshape@port{1}
			}%
		}{}%
		\pgfutil@ifundefined{pgf@anchor@pdshape@out}{
			\expandafter\xdef\csname pgf@anchor@pdshape@out\endcsname{%
				\noexpand\pdshape@port{1}
			}%
		}{}%
	}
}

\def\pdshape@port#1{
	\northeast	

	\ifnum#1=0	
		\if\direction\direce
			\pgf@x=-\pgf@x
			\pgf@ya= \pgf@y
			\pgfmathsetlength{\pgf@y}{\pgf@ya-0.5*\minheight}%
		\fi
		\if\direction\direcw
			\pgf@x=\pgf@x
			\pgf@ya= \pgf@y
			\pgfmathsetlength{\pgf@y}{\pgf@ya-0.5*\minheight}%
		\fi
		\if\direction\direcn
			\pgf@y=-\pgf@y
			\pgf@xa=\pgf@x
			\pgfmathsetlength{\pgf@x}{\pgf@xa-0.5*\minwidth}%
		\fi
		\if\direction\direcs
			\pgf@y=\pgf@y
			\pgf@xa= \pgf@x
			\pgfmathsetlength{\pgf@x}{\pgf@xa-0.5*\minwidth}%
		\fi
	\else	
		\if\direction\direce
			\pgf@x=\pgf@x
			\pgf@ya= \pgf@y
			\pgfmathsetlength{\pgf@y}{\pgf@ya-0.5*\minheight}%
		\fi
		\if\direction\direcw
			\pgf@x=-\pgf@x
			\pgf@ya= \pgf@y
			\pgfmathsetlength{\pgf@y}{\pgf@ya-0.5*\minheight}%
		\fi
		\if\direction\direcn
			\pgf@y=\pgf@y
			\pgf@xa= \pgf@x
			\pgfmathsetlength{\pgf@x}{\pgf@xa-0.5*\minwidth}%
		\fi
		\if\direction\direcs
			\pgf@y=-\pgf@y
			\pgf@xa= \pgf@x
			\pgfmathsetlength{\pgf@x}{\pgf@xa-0.5*\minwidth}%
		\fi
	\fi
}

\pgfaddtoshape{pdshape}{
	\anchorlet{c}{center}
	\anchorlet{n}{north}
	\anchorlet{e}{east}
	\anchorlet{s}{south}
	\anchorlet{w}{west}
	\anchorlet{se}{south east}
	\anchorlet{sw}{south west}
	\anchorlet{ne}{north east}
	\anchorlet{nw}{north west}
	\anchorlet{wsw}{west south west}
	\anchorlet{wnw}{west north west}
	\anchorlet{ene}{east north east}
	\anchorlet{ese}{east south east}
	\anchorlet{nnw}{north north west}
	\anchorlet{nne}{north north east}
	\anchorlet{ssw}{south south west}
	\anchorlet{sse}{south south east}
}

\tikzset{
/tikz/pdkeys/.cd,
size/.initial=0.5,
color/.initial=EO,
direction/.initial=e,
linestyle/.initial={linestyle},
fillgradient/.initial=O,
/tikz/pd/.code={
		\pgfqkeys{/tikz/pdkeys}{#1}%
		\tikzset{/tikz/pdkeys/drawer/.expanded=%
				{\pgfkeysvalueof{/tikz/pdkeys/direction}}%
				{\pgfkeysvalueof{/tikz/pdkeys/size}}%
				{\pgfkeysvalueof{/tikz/pdkeys/color}}%
				{\pgfkeysvalueof{/tikz/pdkeys/linestyle}}%
				{\pgfkeysvalueof{/tikz/pdkeys/fillgradient}}%
		}
	},
/tikz/pdkeys/drawer/.code n args={5}{%
\tikzset{
pdshape,
minimum height=#2*\NODESIZE,
minimum width=#2*\NODESIZE,
#3,
#4,
draw,
append after command={
\pgfextra{\let\bdr=\tikzlastnode%
\node[#5, fit=(\bdr.nw)(\bdr.se)] (boxgradient){};
\draw[---,#3] ($(\bdr.s)!.1!(\bdr.n)$) to ($(\bdr.s)!.9!(\bdr.n)$);
\fill[#3] ({$(\bdr.s)!.3!(\bdr.n)$} -| {$(\bdr.w)!.3!(\bdr.e)$}) to ($(\bdr.s)!.7!(\bdr.n)$) to ({$(\bdr.s)!.3!(\bdr.n)$} -| {$(\bdr.w)!.7!(\bdr.e)$}) to cycle;
\draw[---,#3] ({$(\bdr.s)!.7!(\bdr.n)$} -| {$(\bdr.w)!.35!(\bdr.e)$}) to ({$(\bdr.s)!.7!(\bdr.n)$} -| {$(\bdr.w)!.65!(\bdr.e)$});
}
}
}
},
}
\pgfdeclareshape{pbsshape}{
	\savedmacro\direction{
		\edef\direction{\pgfkeysvalueof{/tikz/pbskeys/direction}}%
	}
	\saveddimen\minwidth{
		\pgfmathsetlength\pgf@x{\pgfshapeminwidth}%
	}
	\saveddimen\minheight{
		\pgfmathsetlength\pgf@x{\pgfshapeminheight}%
	}
	\savedmacro\nport{
		\edef\nport{\pgfkeysvalueof{/tikz/pbskeys/nport}}%
	}
	\inheritsavedanchors[from=basic]

	\inheritanchor[from=basic]{center}
	\inheritanchor[from=basic]{mid}
	\inheritanchor[from=basic]{base}
	\inheritanchor[from=basic]{north}
	\inheritanchor[from=basic]{south}
	\inheritanchor[from=basic]{west}
	\inheritanchor[from=basic]{mid west}
	\inheritanchor[from=basic]{base west}
	\inheritanchor[from=basic]{north west}
	\inheritanchor[from=basic]{south west}
	\inheritanchor[from=basic]{east}
	\inheritanchor[from=basic]{mid east}
	\inheritanchor[from=basic]{base east}
	\inheritanchor[from=basic]{north east}
	\inheritanchor[from=basic]{south east}
	\inheritanchor[from=basic]{west south west}%
	\inheritanchor[from=basic]{west north west}%
	\inheritanchor[from=basic]{east north east}%
	\inheritanchor[from=basic]{east south east}%
	\inheritanchor[from=basic]{north north west}%
	\inheritanchor[from=basic]{north north east}%
	\inheritanchor[from=basic]{south south west}%
	\inheritanchor[from=basic]{south south east}%

	\inheritanchorborder[from=basic]
	\inheritbackgroundpath[from=basic]

	\pgfutil@g@addto@macro\pgf@sh@s@pbsshape{%
		\pgfutil@ifundefined{pgf@anchor@pbsshape@in0}{
			\expandafter\xdef\csname pgf@anchor@pbsshape@in0\endcsname{%
				\noexpand\pbsshape@port[0]{0}
			}%
		}{}%
		\pgfutil@ifundefined{pgf@anchor@pbsshape@in1}{
			\expandafter\xdef\csname pgf@anchor@pbsshape@in1\endcsname{%
				\noexpand\pbsshape@port[1]{0}
			}%
		}{}%
		\pgfutil@ifundefined{pgf@anchor@pbsshape@in}{
			\expandafter\xdef\csname pgf@anchor@pbsshape@in\endcsname{%
				\noexpand\pbsshape@port[0]{0}
			}%
		}{}%
		\pgfutil@ifundefined{pgf@anchor@pbsshape@out0}{
			\expandafter\xdef\csname pgf@anchor@pbsshape@out0\endcsname{%
				\noexpand\pbsshape@port[0]{1}
			}%
		}{}%
		\pgfutil@ifundefined{pgf@anchor@pbsshape@out1}{
			\expandafter\xdef\csname pgf@anchor@pbsshape@out1\endcsname{%
				\noexpand\pbsshape@port[1]{1}
			}%
		}{}%
		\pgfutil@ifundefined{pgf@anchor@pbsshape@out}{
			\expandafter\xdef\csname pgf@anchor@pbsshape@out\endcsname{%
				\noexpand\pbsshape@port[0]{1}
			}%
		}{}%
		\pgfmathsetcount{\portcount}{0}
		\pgfmathloop%
		\ifnum\the\portcount<\nport
		\pgfutil@ifundefined{pgf@anchor@pbsshape@p\the\portcount}{
			\expandafter\xdef\csname pgf@anchor@pbsshape@p\the\portcount\endcsname{%
				\noexpand\pbsshape@port[\the\portcount]{2}
			}%
		}{}%
		\pgfmathaddtocount{\portcount}{1}	
		\repeatpgfmathloop
	}
}

\def\pbsshape@port[#1]#2{
	\northeast	

	\ifnum#2=0	
		\if\direction\direce
			\ifnum#1=0
				\pgf@x=-\pgf@x
				\pgf@ya= \pgf@y
				\pgfmathsetlength{\pgf@y}{\pgf@ya-0.5*\minheight}%
			\else
				\pgf@x=0\pgf@x
				\pgf@ya= \pgf@y
				\pgfmathsetlength{\pgf@y}{\pgf@ya-\minheight}%
			\fi
		\fi
		\if\direction\direcw
			\ifnum#1=0
				\pgf@x=\pgf@x
				\pgf@ya= \pgf@y
				\pgfmathsetlength{\pgf@y}{\pgf@ya-0.5*\minheight}%
			\else
				\pgf@x=0\pgf@x
				\pgf@ya= \pgf@y
				\pgfmathsetlength{\pgf@y}{\pgf@ya}%
			\fi
		\fi
		\if\direction\direcn
			\ifnum#1=0
				\pgf@y=-\pgf@y
				\pgf@xa=\pgf@x
				\pgfmathsetlength{\pgf@x}{\pgf@xa-0.5*\minwidth}%
			\else
				\pgf@y=0\pgf@y
				\pgf@xa=\pgf@x
				\pgfmathsetlength{\pgf@x}{\pgf@xa-0*\minwidth}%
			\fi
		\fi
		\if\direction\direcs
			\ifnum#1=0
				\pgf@y=\pgf@y
				\pgf@xa= \pgf@x
				\pgfmathsetlength{\pgf@x}{\pgf@xa-0.5*\minwidth}%
			\else
				\pgf@y=0\pgf@y
				\pgf@xa= \pgf@x
				\pgfmathsetlength{\pgf@x}{\pgf@xa-1*\minwidth}%
			\fi
		\fi
	\else	
		\if\direction\direce
			\ifnum#1=0
				\pgf@x=\pgf@x
				\pgf@ya= \pgf@y
				\pgfmathsetlength{\pgf@y}{\pgf@ya-0.5*\minheight}%
			\else
				\pgf@x=0\pgf@x
				\pgf@ya= \pgf@y
				\pgfmathsetlength{\pgf@y}{\pgf@ya}%
			\fi
		\fi
		\if\direction\direcw
			\ifnum#1=0
				\pgf@x=-\pgf@x
				\pgf@ya= \pgf@y
				\pgfmathsetlength{\pgf@y}{\pgf@ya-0.5*\minheight}%
			\else
				\pgf@x=0\pgf@x
				\pgf@ya= \pgf@y
				\pgfmathsetlength{\pgf@y}{\pgf@ya-\minheight}%
			\fi
		\fi
		\if\direction\direcn
			\ifnum#1=0
				\pgf@y=\pgf@y
				\pgf@xa= \pgf@x
				\pgfmathsetlength{\pgf@x}{\pgf@xa-0.5*\minwidth}%
			\else
				\pgf@y=0\pgf@y
				\pgf@xa= \pgf@x
				\pgfmathsetlength{\pgf@x}{\pgf@xa-1*\minwidth}%
			\fi
		\fi
		\if\direction\direcs
			\ifnum#1=0
				\pgf@y=-\pgf@y
				\pgf@xa= \pgf@x
				\pgfmathsetlength{\pgf@x}{\pgf@xa-0.5*\minwidth}%
			\else
				\pgf@y=0\pgf@y
				\pgf@xa= \pgf@x
				\pgfmathsetlength{\pgf@x}{\pgf@xa-0*\minwidth}%
			\fi
		\fi
	\fi

	\ifnum#2=2	%
		\northeast \pgf@xa=\pgf@x \pgf@ya=\pgf@y
		\southwest \pgf@xb=\pgf@x \pgf@yb=\pgf@y

		\pgf@xc=\pgf@xa \advance\pgf@xc by -\pgf@xb	
		\pgfmathsetlength{\pgf@x}{\pgf@xa-(#1 + 0.5)*(\pgf@xc/\nport)}%
		\pgf@yc=\pgf@ya \advance\pgf@yc by -\pgf@yb	
		\pgfmathsetlength{\pgf@y}{\pgf@ya-(#1 + 0.5)*(\pgf@yc/\nport)}%

		\if\direction\direce
			\pgf@x=-\pgf@x
		\fi
		\if\direction\direcw
			\pgf@x=-\pgf@x
		\fi
	\fi
}

\pgfaddtoshape{pbsshape}{
	\anchorlet{c}{center}
	\anchorlet{n}{north}
	\anchorlet{e}{east}
	\anchorlet{s}{south}
	\anchorlet{w}{west}
	\anchorlet{se}{south east}
	\anchorlet{sw}{south west}
	\anchorlet{ne}{north east}
	\anchorlet{nw}{north west}
	\anchorlet{wsw}{west south west}
	\anchorlet{wnw}{west north west}
	\anchorlet{ene}{east north east}
	\anchorlet{ese}{east south east}
	\anchorlet{nnw}{north north west}
	\anchorlet{nne}{north north east}
	\anchorlet{ssw}{south south west}
	\anchorlet{sse}{south south east}
}

\pgfmathsetmacro{\WSSSINEHEIGHT}{0.06}

\tikzset{
	/tikz/pbskeys/.cd,
	size/.initial=0.5,
	color/.initial=O,
	direction/.initial=e,
	linestyle/.initial={linestyle},
	nport/.initial=1,
	fillgradient/.initial=O,
	/tikz/pbs/.code={
			\pgfqkeys{/tikz/pbskeys}{#1}%
			\tikzset{/tikz/pbskeys/drawer/.expanded=%
					{\pgfkeysvalueof{/tikz/pbskeys/direction}}%
					{\pgfkeysvalueof{/tikz/pbskeys/size}}%
					{\pgfkeysvalueof{/tikz/pbskeys/color}}%
					{\pgfkeysvalueof{/tikz/pbskeys/linestyle}}%
					{\pgfkeysvalueof{/tikz/pbskeys/fillgradient}}%
			}
		},
	/tikz/pbskeys/drawer/.code n args={5}{%
			\tikzset{
				pbsshape,
				minimum height=#2*\NODESIZE,
				minimum width=#2*\NODESIZE,
				#3,
				#4,
				draw,
				append after command={
						\pgfextra{\let\bdr=\tikzlastnode%
							\node[#5, fit=(\bdr.nw)(\bdr.se)] (boxgradient){};

							\if#1e
								\draw[#3, ---] ($(\bdr.nw)!.01!(\bdr.se)$) to ($(\bdr.se)!.01!(\bdr.nw)$);
							\fi
							\if#1w
								\draw[#3, ---] ($(\bdr.nw)!.01!(\bdr.se)$) to ($(\bdr.se)!.01!(\bdr.nw)$);
							\fi
							\if#1n
								\draw[#3, ---] ($(\bdr.ne)!.01!(\bdr.sw)$) to ($(\bdr.sw)!.01!(\bdr.ne)$);
							\fi
							\if#1s
								\draw[#3, ---] ($(\bdr.ne)!.01!(\bdr.sw)$) to ($(\bdr.sw)!.01!(\bdr.ne)$);
							\fi

						}
					}
			}
		},
}

\ifdefined\fontchoice
\else
	\def\fontchoice{times} 
\fi

\usepackage{pdftexcmds}

\ifnum\pdf@strcmp{\fontchoice}{firasans}=0 %
	\usepackage[book]{FiraSans} 
	\usepackage[T1]{fontenc}
	
	\tikzset{every picture/.style={/utils/exec={\sffamily}}}
\fi

\ifnum\pdf@strcmp{\fontchoice}{times}=0 %
	\usepackage{times}
\fi

\ifnum\pdf@strcmp{\fontchoice}{timesnewroman}=0 %
	\usepackage{mathptmx}
	\usepackage[T1]{fontenc}
\fi

\ifnum\pdf@strcmp{\fontchoice}{helvetica}=0 %
	\usepackage[scaled]{helvet}
	
	\usepackage[T1]{fontenc}
\fi

\makeatother
\pgfplotscreateplotcyclelist{foo}{
        {C1,mark=*},
        {C2,mark=square*},
        {C3, mark = triangle*},
        {C4,mark=+},
        {C5, mark = diamond*},
        {C6, mark = x}
      }
      
\titlespacing{\section}{0pt}{10pt}{5pt}
\usepackage{amsmath,amssymb}
\newcommand{\SetCapsType}{normalcaps}
\usepackage[acronym,nomain]{glossaries}
\glsdisablehyper
\usepackage{xspace}  
\usepackage[shortcuts]{extdash}
\usepackage{listofitems,pgffor}    
\usepackage{siunitx}
\usepackage{xstring}    
\usepackage{silence}
\usepackage{xparse}

\setsepchar{;}  

\ifdefined\silencecommonwarnings
\else
	\def\silencecommonwarnings{true} 
\fi

\ifbool{\silencecommonwarnings}{%
    \WarningFilter{ECOtools}{Cannot define: DH}%
    \WarningFilter{ECOtools}{Cannot define: PAM}%
    \WarningFilter{ECOtools}{Cannot define: QAM}%
    \WarningFilter{ECOtools}{Cannot define: SI}%
    \WarningFilter{ECOtools}{Cannot define: PV}%
    \WarningFilter{ECOtools}{Cannot define: LP}%
    \WarningFilter{ECOtools}{Cannot define: RN}%
    \WarningFilter{ECOtools}{Cannot define: uLP}%
    \WarningFilter{ECOtools}{Redefining DH}%
    }{}

\makeatletter
\providecommand{\SetCapsType}{smallcaps}

\long\def\@scTrue{smallcaps}
\long\def\@scFalse{normalcaps}
\newcommand{\acroSCaps}[1]{%
    \ifx\SetCapsType\@scTrue 
        \textsc{#1}%
    \else
        \MakeUppercase{#1}%
    \fi
}
\makeatother

\usepackage{scalerel}
\makeatletter
\newcommand\scslash{%
\ifx\SetCapsType\@scTrue 
    \protect\stretchrel*{$/$}{\textsc{e}}
\else
    /
\fi
} 
\makeatother 

\makeatletter
\@ifpackageloaded{babel}{%
    \newcommand{\usuk}[2]{%
        \iflanguage{USenglish}{#1}{#2}%
    }%
}{%
    \newcommand{\usuk}[2]{%
        #1%
    }%
}%

\newcommand{\langcheck}[2]{
    \@ifpackageloaded{babel}{%
        \iflanguage{USenglish}{#1}{#2}%
    }{%
        #1%
    }%
}

\makeatother

\newcommand{\short}[1]{%
    \glsentrytext{#1}\xspace%
}
\newcommand{\shortfakeplural}[1]{%
    \glsentrytext{#1}s\xspace%
}
\newcommand{\Short}[1]{%
    \Glsentrytext{#1}\xspace%
}
\newcommand{\normal}[1]{%
    \gls{#1}\xspace%
}
\newcommand{\longacr}[1]{%
    \acrlong{#1}\xspace%
}
\newcommand{\plural}[1]{%
    \glspl{#1}\xspace%
}
\newcommand{\full}[1]{%
    \acrfull{#1}\xspace%
}
\newcommand{\fullplural}[1]{%
    \acrfullpl{#1}\xspace%
}
\newcommand{\Normal}[1]{%
    \Gls{#1}\xspace%
}
\newcommand{\Plural}[1]{%
    \Glspl{#1}\xspace%
}
\newcommand{\Full}[1]{%
    \Acrfull{#1}\xspace%
}
\newcommand{\Fullplural}[1]{%
    \Acrfullpl{#1}\xspace%
} 

\newcommand{\texpdfif}[2]{%
    \ifcsname texorpdfstring\endcsname%
        \texorpdfstring{#1{#2}}{#2\xspace}%
    \else%
        #1{#2}%
    \fi%
}

\newcommand{\checkanddefine}[3]{%
	\ifcsname #1\endcsname%
        \PackageWarning{ECOtools}{Cannot define: #1 already defined, trying to define g#1 instead.}%
        \ifcsname g#1\endcsname%
            \PackageWarning{ECOtools}{Cannot define: g#1 also already defined.}%
    	\else%
        	\expandafter\newcommand\csname g#1\endcsname{%
        	    \texpdfif{#2}{#3}%
    	    }%
        \fi%
	\else%
    	\expandafter\newcommand\csname #1\endcsname{%
    	    \texpdfif{#2}{#3}%
	    }%
    \fi%
}

\newcommand{\redefine}[3]{%
    \PackageWarning{ECOtools}{Redefining #1}%
	\expandafter\renewcommand\csname #1\endcsname{%
	    \texpdfif{#2}{#3}%
    }%
}

\newcommand{\nAcronym}[4][]{%
	\newacronym[#1]{#2}{#3}{#4}%
	\checkanddefine{s#2}{\short}{#2}%
    \checkanddefine{s#2s}{\shortfakeplural}{#2}%
	\checkanddefine{#2}{\normal}{#2}%
	\checkanddefine{l#2}{\longacr}{#2}%
	\checkanddefine{#2s}{\plural}{#2}%
	\checkanddefine{f#2}{\full}{#2}%
	\checkanddefine{f#2s}{\fullplural}{#2}%
	\checkanddefine{su#2}{\Short}{#2}%
	\checkanddefine{u#2}{\Normal}{#2}%
	\checkanddefine{u#2s}{\Plural}{#2}%
	\checkanddefine{fu#2}{\Full}{#2}%
	\checkanddefine{fu#2s}{\Fullplural}{#2}%
	\IfStrEq{#2}{DH}{
	    \redefine{#2}{\normal}{#2}%
	    }{}%
}%

\NewDocumentCommand\qam{g}{%
    \IfNoValueTF{#1}{%
        \texpdfif{\gls}{QAM}\xspace%
        }{%
        \StrLen{#1}[\stringlength]%
        \ifnum\stringlength=0%
            \texpdfif{\gls}{QAM}\xspace%
        \else%
            {\qamlisthelper{#1}}%
        \fi%
        }%
}

\let\QAM\qam

\DeclareRobustCommand\qamlisthelper[1]{%
    \readlist*\args{#1}%
    \acroSCaps{\args[1]\=/}%
    \ifnum\argslen = 2%
        { and \acroSCaps{\args[2]}\=/}%
    \fi%
    \ifnum\argslen > 2%
        \foreach \n in {2,...,\argslen}{%
            \ifnum\n = \argslen%
                {, and }%
            \else 
                {, }%
            \fi%
            {\acroSCaps{\args[\n]}\=/}%
        }%
    \fi%
    \ifglsused{QAM}%
        {}%
        {ary }%
    \texpdfif{\gls}{QAM}%
}%

\NewDocumentCommand\pam{g}{%
    \IfNoValueTF{#1}{%
        \texpdfif{\gls}{PAM}\xspace%
        }{%
        \StrLen{#1}[\stringlength]%
        \ifnum\stringlength=0%
            \texpdfif{\gls}{PAM}\xspace%
        \else%
            {\pamlisthelper{#1}}%
        \fi%
        }%
}

\DeclareRobustCommand\pamlisthelper[1]{%
    \readlist*\args{#1}%
    \ifglsused{PAM}{%
        \texpdfif{\gls}{PAM}%
        \acroSCaps{\=/\args[1]}%
        \ifnum\argslen = 2%
            { and \=/\acroSCaps{\args[2]}}%
        \fi%
        \ifnum\argslen > 2%
            \foreach \n in {2,...,\argslen}{%
                \ifnum\n = \argslen%
                    {, and }%
                \else%
                    {, }%
                \fi%
                {\=/\acroSCaps{\args[\n]}}%
            }%
        \fi%
    }{%
        \acroSCaps{\args[1]\=/}%
        \ifnum\argslen = 2%
            { and \acroSCaps{\args[2]}\=/}%
        \fi%
        \ifnum\argslen > 2%
            \foreach \n in {2,...,\argslen}{%
                \ifnum\n = \argslen%
                    {, and }%
                \else%
                    {, }%
                \fi
                {\acroSCaps{\args[\n]}\=/}%
            }%
        \fi%
        {ary }%
        \texpdfif{\gls}{PAM}%
    }%
}%

\NewDocumentCommand\lp{g}{%
    \IfNoValueTF{#1}{%
        \texpdfif{\normal}{LP}%
        }{%
        \StrLen{#1}[\stringlength]%
        \ifnum\stringlength=0%
            \texpdfif{\normal}{LP}%
        \else%
            \ifglsused{LP}{}{\texpdfif{\normal}{LP}\xspace}%
            \lplisthelper[lp]{#1}%
        \fi%
        }%
}

\NewDocumentCommand\ulp{g}{%
    \IfNoValueTF{#1}{%
        \texpdfif{\Normal}{LP}\xspace%
        }{%
        \StrLen{#1}[\stringlength]%
        \ifnum\stringlength=0%
            \texpdfif{\Normal}{LP}\xspace%
        \else%
            \ifglsused{LP}{%
                \lplisthelper[Lp]{#1}%
            }{%
                \texpdfif{\Normal}{LP}\xspace\lplisthelper[lp]{#1}%
            }%
        \fi%
        }%
}
\DeclareRobustCommand\lplisthelper[2][lp]{%
    \readlist*\args{#2}%
    \foreach \n in {1,...,\argslen}{%
        \ifnum \n > 1%
            \ifnum \argslen > 2%
                {, }%
            \else%
                { }%
            \fi%
        \fi%
        \ifnum \n = \argslen%
            \ifnum \argslen > 1%
                {and }%
            \fi%
        \fi%
        \ifnum \n = 1%
            {\acroSCaps{#1}}
        \else%
            {\acroSCaps{\MakeLowercase{#1}}}%
        \fi%
        {\textsubscript{\StrSplit{\args[\n]}{2}{\csA}{\csB}\acroSCaps{\csA}\csB}}
    }%
}%

\nAcronym{128SPQAM}{\acroSCaps{128-sp-16-qam}}{128-ary set-partitioning \QAM{16}}

\nAcronym{2A8PSK}{\acroSCaps{2a8psk}}{2-ary amplitude 8-ary phaseshift keying}

\nAcronym{3CCMCF}{\acroSCaps{3cc-mcf}}{3-core coupled-core multi-core fiber}

\nAcronym{4D}{\acroSCaps{4d}}{four-dimensional}
\nAcronym{4D64PRS}{\acroSCaps{4d-64prs}}{\usuk{four-dimensional 64-ary polarization-ring-switching}{four-dimensional 64-ary polarisation-ring-switching}}
\nAcronym{4DOS128}{\acroSCaps{4d-os128}}{four-dimensional orthant-symmetric 128-ary modulation format}

\nAcronym{5B4D2A8PSK}{\acroSCaps{5b4d-2a8psk}}{5-bit four-dimensional two-amplitude 8-ary phase-shift keying}

\nAcronym{6B4D2A8PSK}{\acroSCaps{6b4d-2a8psk}}{6-bit four-dimensional two-amplitude 8-ary phase-shift keying}

\nAcronym{7B4D2A8PSK}{\acroSCaps{7b4d-2a8psk}}{7-bit four-dimensional two-amplitude 8-ary phase-shift keying}

\nAcronym{8D}{\acroSCaps{8d}}{eight-dimensional}\nAcronym{8D2048PRS}{\acroSCaps{8d-2048prs}}{eight-dimensional 2048-ary polarization-ring-switching}
\nAcronym{8D2048PRST1}{\acroSCaps{8d-2048prs-t1}}{eight-dimensional 2048-ary polarization-ring-switching type 1}
\nAcronym{8D2048PRST2}{\acroSCaps{8d-2048prs-t2}}{eight-dimensional 2048-ary polarization-ring-switching type 2}
\nAcronym{8DAPSK}{\acroSCaps{8d-apsk}}{eight-dimensional amplitude-phase-shift keying}

\nAcronym{ABC}{\acroSCaps{abc}}{automatic bias control}
\nAcronym{AC}{\acroSCaps{ac}}{alternating current}
\nAcronym{ADC}{\acroSCaps{adc}}{\usuk{analog-to-digital converter}{analogue-to-digital converter}}
\nAcronym{AGC}{\acroSCaps{agc}}{automatic gain control}
\nAcronym{AIR}{\acroSCaps{air}}{achievable information rate}
\nAcronym{AMZI}{\acroSCaps{amzi}}{asymmetric Mach–Zehnder interferometer}
\nAcronym{AO}{\acroSCaps{ao}}{adaptive optics}
\nAcronym{AOM}{\acroSCaps{aom}}{acousto-optic modulator}
\nAcronym{APD}{\acroSCaps{apd}}{avalanche photodiode}
\nAcronym{API}{\acroSCaps{api}}{application programming interface}
\nAcronym{AR}{\acroSCaps{ar}}{achievable rate}
\nAcronym{ARRWG}{\acroSCaps{a}rr\acroSCaps{wg}}{arrayed-waveguide grating}
\nAcronym{ASE}{\acroSCaps{ase}}{amplified spontaneous emission}
\nAcronym{ASK}{\acroSCaps{ask}}{amplitude-shift keying}
\nAcronym{ASIC}{\acroSCaps{asic}}{application-specific integrated circuit}
\nAcronym{ATS}{\acroSCaps{ats}}{alignment tracking sensor}
\nAcronym{AWG}{\acroSCaps{awg}}{arbitrary-waveform generator}
\nAcronym{AWGN}{\acroSCaps{awgn}}{additive white Gaussian noise}

\nAcronym{BBU}{\acroSCaps{bbu}}{baseband unit}
\nAcronym{BCH}{\acroSCaps{bch}}{Bose-Chaudhuri-Hocquenghem}
\nAcronym{BER}{\acroSCaps{ber}}{bit error rate}
\nAcronym{BERT}{\acroSCaps{bert}}{bit error rate tester}
\nAcronym{BICM}{\acroSCaps{bicm}}{bit-interleaved coded modulation}
\nAcronym{BMD}{\acroSCaps{bmd}}{bit-metric decoding}
\nAcronym{BPD}{\acroSCaps{bpd}}{balanced photo-diode}
\nAcronym{BPF}{\acroSCaps{bpf}}{bandpass filter}
\nAcronym{BPS}{\acroSCaps{bps}}{blind phase search}
\nAcronym{BPSK}{\acroSCaps{bpsk}}{binary phase-shift keying}
\nAcronym{BRGC}{\acroSCaps{brgc}}{binary reflected Gray code}
\nAcronym{BTB}{\acroSCaps{btb}}{back-to-back}

\nAcronym{CAGR}{\acroSCaps{cagr}}{compound annual growth rate}
\nAcronym{CCDM}{\acroSCaps{ccdm}}{constant composition distribution matching}
\langcheck{%
    \nAcronym{CCF}{\acroSCaps{ccf}}{coupled-core fiber}%
    }{%
    \nAcronym{CCF}{\acroSCaps{ccf}}{coupled-core fibre}%
}%
\nAcronym{CD}{\acroSCaps{cd}}{chromatic dispersion}
\nAcronym{CIR}{\acroSCaps{cir}}{channel impulse response}
\nAcronym{CMA}{\acroSCaps{cma}}{constant modulus algorithm}
\nAcronym{CMF}{\acroSCaps{cmf}}{core multiplicity factor}
\nAcronym{CMUX}{\acroSCaps{cmux}}{core multiplexer}
\nAcronym{COTS}{\acroSCaps{cots}}{commercial off-the-shelf}
\nAcronym{COW}{\acroSCaps{cow}}{coherent one-way}
\nAcronym{ChUT}{\acroSCaps{chut}}{channel under test}
\nAcronym[firstplural=channels under test (\acroSCaps{cut}s)]{CUT}{\acroSCaps{cut}}{channel under test}
\nAcronym{CRX}{\acroSCaps{crx}}{coherent receiver}
\nAcronym{CPE}{\acroSCaps{cpe}}{carrier phase estimation}
\nAcronym{CPU}{\acroSCaps{cpu}}{central processing unit}
\nAcronym{CSPR}{\acroSCaps{cspr}}{carrier-to-signal power ratio}
\nAcronym{CUDA}{\acroSCaps{cuda}}{compute unified device architecture}
\nAcronym{CVQKD}{\acroSCaps{cv-qkd}}{continuous-variable quantum key distribution}
\nAcronym{CW}{\acroSCaps{cw}}{continuous wave}
\nAcronym{CCD}{\acroSCaps{ccd}}{charge-coupled device}

\nAcronym{DA}{\acroSCaps{da}}{driver amplifier}
\nAcronym{DAC}{\acroSCaps{dac}}{\usuk{digital-to-analog converter}{digital-to-analogue converter}}
\nAcronym{DC}{\acroSCaps{dc}}{direct current}
\nAcronym{DBP}{\acroSCaps{dbp}}{digital backpropagation}
\nAcronym{DCF}{\acroSCaps{dcf}}{\usuk{dispersion-compensating fiber}{dispersion-compensating fibre}}
\langcheck{%
    \nAcronym{DCI}{\acroSCaps{dci}}{data center interconnect}
    }{%
    \nAcronym{DCI}{\acroSCaps{dci}}{data centre interconnect}
}%
\nAcronym{DDLMS}{\acroSCaps{dd-lms}}{decision-directed least mean square}
\nAcronym{DEMUX}{\acroSCaps{demux}}{de-multiplexer}
\nAcronym{DFA}{\acroSCaps{dfa}}{\usuk{doped fiber amplifier}{doped fibre amplifier}}
\nAcronym{DFB}{\acroSCaps{dfb}}{distributed feedback}
\nAcronym{DGD}{\acroSCaps{dgd}}{differential group delay}
\nAcronym{DH}{\acroSCaps{dh}}{digital holography}
\nAcronym{DM}{\acroSCaps{dm}}{distribution matcher}
\nAcronym{DMR}{\acroSCaps{dm}}{dichroic mirror}
\nAcronym{DMA}{\acroSCaps{dma}}{direct memory access}
\nAcronym{DMD}{\acroSCaps{dmd}}{differential mode delay}
\nAcronym{DMG}{\acroSCaps{dmg}}{differential modal gain}
\nAcronym{DMGD}{\acroSCaps{dmgd}}{differential mode group delay}
\nAcronym{DML}{\acroSCaps{dml}}{directly-modulated laser}
\nAcronym{DP}{\acroSCaps{dp}}{\usuk{dual-polarization}{dual-polarisation}}
\nAcronym{DPC}{\acroSCaps{dpc}}{digital pre-compensation}
\nAcronym{DPE}{\acroSCaps{dpe}}{digital pre-emphasis}
\nAcronym{DPIQ}{\acroSCaps{dp-iqm}}{\usuk{dual-polarization \acroSCaps{iq}-modulator}{dual-polarisation \acroSCaps{iq}-modulator}}
\nAcronym{DPLL}{\acroSCaps{dpll}}{digital phase-locked loop}
\nAcronym{DPS}{\acroSCaps{dps}}{differential phase shift}
\nAcronym{DQPSK}{\acroSCaps{dqpsk}}{differential quaternary phase-shift-keying}
\nAcronym{DRA}{\acroSCaps{dra}}{distributed Raman amplifier}
\nAcronym{DRE}{\acroSCaps{dre}}{digital resolution enhancer}
\nAcronym{DSB}{\acroSCaps{dsb}}{double-sideband}
\nAcronym{DSF}{\acroSCaps{dsf}}{\usuk{dispersion-shifted fiber}{dispersion-shifted fibre}} 
\nAcronym{DSO}{\acroSCaps{dso}}{digital sampling oscilloscope}
\nAcronym{DSP}{\acroSCaps{dsp}}{digital signal processing}
\nAcronym{DUT}{\acroSCaps{dut}}{device-under-test}

\nAcronym{DVQKD}{\acroSCaps{dv-qkd}}{discrete-variable quantum key distribution}

\nAcronym{DWDM}{\acroSCaps{dwdm}}{dense wavelength-division multiplexing}

\nAcronym{EAM}{\acroSCaps{eam}}{electro-absorption modulator}
\nAcronym{ECL}{\acroSCaps{ecl}}{external cavity laser}
\nAcronym{ED}{\acroSCaps{ed}}{Eucledian distance}
\nAcronym{EDF}{\acroSCaps{edf}}{\usuk{erbium-doped fiber}{erbium-doped fibre}}
\nAcronym{EDFA}{\acroSCaps{edfa}}{\usuk{erbium-doped fiber amplifier}{erbium-doped fibre amplifier}}
\nAcronym{ENOB}{\acroSCaps{enob}}{effective number of bits}
\nAcronym{ER}{\acroSCaps{er}}{extinction ratio}
\nAcronym{ESS}{\acroSCaps{ess}}{enumerative sphere shaping}

\langcheck{%
    \nAcronym{FBG}{\acroSCaps{fbg}}{fiber Bragg grating}%
    }{%
    \nAcronym{FBG}{\acroSCaps{fbg}}{fibre Bragg grating}%
}%
\nAcronym{FD}{\acroSCaps{fd}}{frequency domain}
\nAcronym{FDE}{\acroSCaps{fde}}{\usuk{frequency domain equalizer}{frequency domain equaliser}}
\nAcronym{FEC}{\acroSCaps{fec}}{forward error correction}
\nAcronym{FFE}{\acroSCaps{ffe}}{\usuk{feed-forward equalizer}{feed-forward equaliser}}
\nAcronym{FFT}{\acroSCaps{fft}}{fast Fourier transform}
\nAcronym{FIR}{\acroSCaps{fir}}{finite impulse response}
\nAcronym{FLOPS}{\acroSCaps{flops}}{floating point operations per second}
\nAcronym{FMEDF}{\acroSCaps{fm-edf}}{\usuk{few-mode erbium-doped fiber}{few-mode erbium-doped fibre}}
\nAcronym{FMEDFA}{\acroSCaps{fm-edfa}}{\usuk{few-mode erbium-doped fiber amplifier}{few-mode erbium-doped fibre amplifier}}
\langcheck{%
    \nAcronym{FMF}{\acroSCaps{fmf}}{few-mode fiber}%
    }{%
    \nAcronym{FMF}{\acroSCaps{fmf}}{few-mode fibre}%
}%
\nAcronym[plural=FM-MCF, firstplural=\usuk{few-mode multi-core fibers}{few-mode multi-core fibres}]{FMMCF}{\acroSCaps{fm-mcf}}{\usuk{few-mode multi-core fiber}{few-mode multi-core fibre}}
\langcheck{%
    \nAcronym{FMPBGF}{\acroSCaps{fm-pbgf}}{few-mode photonic bandgap fiber}%
    }{%
    \nAcronym{FMPBGF}{\acroSCaps{fm-pbgf}}{few-mode photonic bandgap fibre}%
}%
\nAcronym{FOV}{\acroSCaps{fov}}{field of view}
\nAcronym{FPGA}{\acroSCaps{fpga}}{field-programmable gate array}
\nAcronym{FWM}{\acroSCaps{fwm}}{four-wave mixing}
\nAcronym{FSO}{\acroSCaps{fso}}{free-space optical}
\nAcronym{FUT}{\acroSCaps{fut}}{\usuk{fiber under test}{fibre under test}}

\nAcronym{GD}{\acroSCaps{gd}}{group delay}
\nAcronym{GI}{\acroSCaps{gi}}{graded-index}
\nAcronym{GFF}{\acroSCaps{gff}}{gain flattening filter}
\nAcronym{GIFMF}{\acroSCaps{gi-fmf}}{\usuk{graded-index few-mode fiber}{graded-index few-mode fibre}}
\nAcronym{GIMMF}{\acroSCaps{gi-mmf}}{\usuk{graded-index multi-mode fiber}{graded-index multi-mode fibre}}
\nAcronym{GMI}{\acroSCaps{gmi}}{\usuk{generalized mutual information}{generalised mutual information}}
\nAcronym{GNSE}{\acroSCaps{gnse}}{generalized nonlinear Schr\"{o}dinger equation}
\nAcronym{GPU}{\acroSCaps{gpu}}{graphics processing unit}
\nAcronym{GS}{\acroSCaps{gs}}{geometric shaping}
\nAcronym{GV}{\acroSCaps{gv}}{group velocity}
\nAcronym{GVD}{\acroSCaps{gvd}}{group velocity dispersion}
\nAcronym{GPIO}{\acroSCaps{gpio}}{general purpose input output}
\nAcronym{GUI}{\acroSCaps{gui}}{graphical user interface}

\langcheck{%
    \nAcronym{HCF}{\acroSCaps{hcf}}{hollow-core fiber}
    }{%
    \nAcronym{HCF}{\acroSCaps{hcf}}{hollow-core fibre}
}%
\nAcronym{HDFEC}{\acroSCaps{hd-fec}}{hard-decision forward error correction}
\nAcronym{HG}{\acroSCaps{hg}}{Hermite-Gaussian}
\nAcronym{HOM}{\acroSCaps{hom}}{higher-order modes}
\nAcronym{HV}{\acroSCaps{hv}}{Hufnagel-Valley}
\nAcronym{HAP}{\acroSCaps{hap}}{Hufnagel-Andrew-Phillips}

\nAcronym{ICS}{\acroSCaps{ics}}{inter-core skew}
\nAcronym{ICXT}{\acroSCaps{ic-xt}}{inter-core cross-talk}
\nAcronym[plural=IL, firstplural=insertion losses (\acroSCaps{il})]{IL}{\acroSCaps{il}}{insertion loss}
\nAcronym{IFFT}{\acroSCaps{ifft}}{inverse fast Fourier transform}
\nAcronym{IIR}{\acroSCaps{iir}}{intensity impulse response}
\nAcronym{IM}{\acroSCaps{im}}{intensity modulator}
\nAcronym{IMDD}{\acroSCaps{im}\scslash \acroSCaps{dd}}{intensity-modulation direct-detection}
\nAcronym{IQM}{\acroSCaps{iqm}}{in-phase and quadrature modulator}
\nAcronym{ISI}{\acroSCaps{isi}}{inter-symbol interference}
\nAcronym{IP}{\acroSCaps{ip}}{intellectual property}

\nAcronym{JGN}{\acroSCaps{jgn}}{Japan Gigabit Network}

\nAcronym{KK}{\acroSCaps{kk}}{Kramers-Kronig}
\nAcronym{KIT}{\acroSCaps{kit}}{Karlsruhe Institute of Technology}

\nAcronym{LCOS}{\acroSCaps{LCoS}}{liquid crystal on silicon}
\nAcronym{LDPC}{\acroSCaps{ldpc}}{low-density parity-check}
\nAcronym{LEAF}{\acroSCaps{leaf}}{\usuk{large effective area fiber}{large effective area fibre}}
\nAcronym{LFSR}{\acroSCaps{lfsr}}{linear-feedback shift register}
\nAcronym{LG}{\acroSCaps{lg}}{Laguerre-Gaussian}
\nAcronym{LMS}{\acroSCaps{lms}}{least mean square}
\nAcronym{LLR}{\acroSCaps{llr}}{log-likelihood ratio}
\nAcronym{LO}{\acroSCaps{lo}}{local oscillator}
\nAcronym{LP}{\acroSCaps{lp}}{\usuk{linearly polarized}{linearly polarised}}
\nAcronym{LSPS}{\acroSCaps{lsps}}{\usuk{loop-synchronized polarization scrambler}{loop-synchronised polarisation scrambler}}
\nAcronym{LUT}{\acroSCaps{lut}}{lookup table}

\nAcronym{MVM}{\acroSCaps{mvm}}{matrix-vector multiplication}
\nAcronym{MB}{\acroSCaps{mb}}{Maxwell-Bolzmann}
\langcheck{%
    \nAcronym{MCF}{\acroSCaps{mcf}}{multi-core fiber}%
    }{%
    \nAcronym{MCF}{\acroSCaps{mcf}}{multi-core fibre}%
}%
\nAcronym{MDG}{\acroSCaps{mdg}}{mode dependent gain}
\nAcronym[firstplural=mode-dependent losses (\acroSCaps{mdl})]{MDL}{\acroSCaps{mdl}}{mode-dependent loss}
\nAcronym{MDM}{\acroSCaps{mdm}}{mode-division multiplexing}
\nAcronym{MEMS}{\acroSCaps{mems}}{micro-electro-mechanical systems}
\nAcronym{MF}{\acroSCaps{mf}}{matched filter}
\nAcronym{MFD}{\acroSCaps{mfd}}{mode field diameter}
\nAcronym{MI}{\acroSCaps{mi}}{mutual information}
\nAcronym{MIMO}{\acroSCaps{mimo}}{multiple-input multiple-output}
\nAcronym{ML}{\acroSCaps{ml}}{machine learning}
\nAcronym{MMA}{\acroSCaps{mma}}{multi-modulus algorithm}
\nAcronym{MMEDF}{\acroSCaps{mmedf}}{\usuk{multi-mode erbium-doped fiber}{multi-mode erbium-doped fibre}}
\nAcronym{MMEDFA}{\acroSCaps{mmedfa}}{\usuk{multi-mode erbium-doped fiber amplifier}{multi-mode erbium-doped fibre amplifier}}
\nAcronym{MMF}{\acroSCaps{mmf}}{\usuk{multi-mode fiber}{multi-mode fibre}}
\nAcronym{MMSE}{\acroSCaps{mmse}}{minimum mean squared error}
\nAcronym{MP}{\acroSCaps{mp}}{minimum phase}
\nAcronym{MPLC}{\acroSCaps{mplc}}{multi-plane light converter}
\nAcronym{MRC}{\acroSCaps{mrc}}{maximum ratio combining}
\nAcronym{MSE}{\acroSCaps{mse}}{mean squared error}
\nAcronym{MUX}{\acroSCaps{mux}}{multiplexer}
\nAcronym{MZM}{\acroSCaps{mzm}}{Mach-Zehnder modulator}
\nAcronym{MZI}{\acroSCaps{mzi}}{Mach-Zehnder interferometer}

\nAcronym{NA}{\acroSCaps{na}}{numerical aperture}
\langcheck{%
    \nAcronym{NANF}{\acroSCaps{nanf}}{nested antiresonant nodeless fiber}%
    }{%
    \nAcronym{NANF}{\acroSCaps{nanf}}{nested antiresonant nodeless fibre}%
}%
\nAcronym{NF}{\acroSCaps{nf}}{noise figure}
\nAcronym{NGMI}{\acroSCaps{ngmi}}{\usuk{normalized generalized mutual information}{normalised generalised mutual information}}
\nAcronym{NLSE}{\acroSCaps{nlse}}{nonlinear Schr\"{o}ding equation}
\nAcronym{NN}{\acroSCaps{nn}}{neural network}
\nAcronym{NIC}{\acroSCaps{nic}}{network interface card}
\nAcronym{NICT}{\acroSCaps{nict}}{National Institute of Information and Communications Technology}
\nAcronym{NIR}{\acroSCaps{nir}}{near-infrared}
\nAcronym{NISTSTS}{\acroSCaps{nist-sts}}{National Insitute of Standards and Technology: Statistical Test Suite}
\nAcronym{NRZ}{\acroSCaps{nrz}}{non-return-to-zero}
\nAcronym{NZDSF}{\acroSCaps{nz-dsf}}{\usuk{non-zero dispersion-shifted fiber}{non-zero dispersion-shifted fibre}} 

\nAcronym{OAM}{\acroSCaps{oam}}{orbital angular momentum}
\nAcronym{OBTB}{\acroSCaps{obtb}}{optical back-to-back}
\nAcronym{OCT}{\acroSCaps{oct}}{outer cladding thickness}
\nAcronym{ODE}{\acroSCaps{ode}}{ordinary differential equation}
\nAcronym{ODL}{\acroSCaps{odl}}{optical delay line}
\nAcronym{OEO}{\acroSCaps{oeo}}{optical-electrical-optical}
\nAcronym{OFC}{\acroSCaps{ofc}}{Optical Fiber Communications Conference}
\nAcronym{OFDR}{\acroSCaps{ofdr}}{optical frequency-domain reflectometer} 
\nAcronym{OFDM}{\acroSCaps{ofdm}}{orthogonal frequency division multiplexing}
\nAcronym{OH}{\acroSCaps{oh}}{overhead}
\nAcronym{OMFT}{\acroSCaps{omft}}{optical multi-format transmitter}
\nAcronym{OOK}{\acroSCaps{ook}}{on-off keying}
\nAcronym{OP}{\acroSCaps{op}}{optical processor}
\nAcronym{OPLL}{\acroSCaps{opll}}{optical phase-locked loop}
\nAcronym{OSA}{\acroSCaps{osa}}{\usuk{optical spectrum analyzer}{optical spectrum analyser}}
\nAcronym{OSNR}{\acroSCaps{osnr}}{optical signal-to-noise ratio}
\nAcronym{OTDR}{\acroSCaps{otdr}}{optical time-domain reflectometer}
\nAcronym{OTF}{\acroSCaps{otf}}{optical tunable filter}
\langcheck{%
    \nAcronym{OVNA}{\acroSCaps{ovna}}{optical vector network analyzer}%
    }{%
    \nAcronym{OVNA}{\acroSCaps{ovna}}{optical vector network analyser}%
}%
\nAcronym{OTG}{\acroSCaps{otg}}{optical turbulence generator}

\nAcronym{PAM}{\acroSCaps{pam}}{pulse-amplitude modulation}
\nAcronym{PAS}{\acroSCaps{pas}}{probabilistic amplitude shaping}
\nAcronym{PAPR}{\acroSCaps{papr}}{peak-to-average power ratio}
\nAcronym{PBC}{\acroSCaps{pbc}}{\usuk{polarization beam combiner}{polarisation beam combiner}}
\langcheck{%
    \nAcronym{PBGF}{\acroSCaps{pbgf}}{photonic bandgap fiber}%
    }{%
    \nAcronym{PBGF}{\acroSCaps{pbgf}}{photonic bandgap fibre}%
}%
\nAcronym{PBS}{\acroSCaps{pbs}}{polarization beam splitter}
\nAcronym{PC}{\acroSCaps{pc}}{physical contact}
\nAcronym{PCVD}{\acroSCaps{pcvd}}{plasma chemical vapor depostion}
\nAcronym{PCG}{\acroSCaps{pcg64}}{64-bit permuted congruential generator}
\nAcronym{PD}{\acroSCaps{pd}}{photodiode}
\nAcronym{PDF}{\acroSCaps{pdf}}{probability density function}
\langcheck{%
    \nAcronym{PDL}{\acroSCaps{pdl}}{polarization-dependent loss}
    }{%
    \nAcronym{PDL}{\acroSCaps{pdl}}{polarisation-dependent loss}
}%
\nAcronym{PDM}{\acroSCaps{pdm}}{\usuk{polarization-division multiplexing}{polarisation-division multiplexing}}
\langcheck{%
    \nAcronym{PER}{\acroSCaps{per}}{polarization extinction ratio}%
    }{%
    \nAcronym{PER}{\acroSCaps{per}}{polarisation extinction ratio}%
}%
\nAcronym{PIC}{\acroSCaps{pic}}{photonic integrated circuit}
\nAcronym{PL}{\acroSCaps{pl}}{photonic lantern}
\nAcronym{PM}{\acroSCaps{pm}}{polarization-multiplexed}
\nAcronym{PMBPSK}{\acroSCaps{pm-bpsk}}{polarization-multiplexed binary phase-shift keying}
\nAcronym{PMQPSK}{\acroSCaps{pm-qpsk}}{polarization-multiplexed quaternary phase-shift keying}
\nAcronym{PM8QAM}{\acroSCaps{pm-8qam}}{polarization-multiplexed 8-ary quadrature amplitude modulation}
\nAcronym{PMD}{\acroSCaps{pmd}}{\usuk{polarization mode dispersion}{polarisation mode dispersion}}
\langcheck{%
    \nAcronym{PMF}{\acroSCaps{pmf}}{polarization-maintaining fiber}%
    }{%
    \nAcronym{PMF}{\acroSCaps{pmf}}{polarization-maintaining fibre}%
}%
\nAcronym{PMP}{\acroSCaps{pmp}}{phase-matching point}
\nAcronym{PNOB}{\acroSCaps{pnob}}{physical number of bits}
\nAcronym{PON}{\acroSCaps{pon}}{passive-optical network}
\nAcronym{PRBS}{\acroSCaps{prbs}}{pseudorandom bit sequence}
\nAcronym{PRNG}{\acroSCaps{PRNG}}{pseudo random number generator}
\nAcronym{PROFA}{\acroSCaps{profa}}{\usuk{pitch reducing optical fiber array}{pitch reducing optical fibre array}}
\nAcronym{PPM}{\acroSCaps{ppm}}{pulse-position modulation}
\nAcronym{PS}{\acroSCaps{ps}}{probabilistic shaping}
\langcheck{%
    \nAcronym{PSCF}{\acroSCaps{pscf}}{pure silica core fiber}%
    }{%
    \nAcronym{PSCF}{\acroSCaps{pscf}}{pure silica core fibre}%
}%
\nAcronym{PSD}{\acroSCaps{psd}}{power spectral density}
\nAcronym{PSF}{\acroSCaps{psf}}{point spread function}
\nAcronym{PSK}{\acroSCaps{psk}}{phase-shift keying}
\nAcronym{PSP}{\acroSCaps{psp}}{\usuk{principal states of polarization}{principal states of polarisation}}
\nAcronym{PSW}{\acroSCaps{psw}}{\usuk{polarization switch}{polarisation switch}}
\nAcronym{PID}{\acroSCaps{pid}}{proportional–integral–derivative}
\nAcronym{PV}{\acroSCaps{pv}}{process value}

\nAcronym{QAM}{\acroSCaps{qam}}{quadrature amplitude modulation}
\nAcronym{QBER}{\acroSCaps{qber}}{quantum bit error rate}
\nAcronym{QKD}{\acroSCaps{qkd}}{quantum key distribution}
\nAcronym{QPSK}{\acroSCaps{qpsk}}{quadrature phase-shift keying}
\nAcronym{QRNG}{\acroSCaps{qrng}}{quantum random number generator}
\nAcronym{QSM}{\acroSCaps{qsm}}{quasi-single-mode}

\nAcronym{RAM}{\acroSCaps{ram}}{random-access memory}
\nAcronym{RC}{\acroSCaps{rc}}{raised cosine}
\nAcronym{RCMF}{\acroSCaps{rcmf}}{relative core multiplicity factor}
\nAcronym{RCMCF}{\acroSCaps{rc-mcf}}{\usuk{randomly-coupled multi-core fiber}{randomly-coupled multi-core fibre}}
\nAcronym{RF}{\acroSCaps{rf}}{radio frequency}
\nAcronym{RFSoC}{\acroSCaps{rfsoc}}{radio frequency system-on-chip}
\nAcronym{RI}{\acroSCaps{ri}}{refractive index}
\nAcronym{RLS}{\acroSCaps{rls}}{recursive least squares}
\nAcronym{RRC}{\acroSCaps{rrc}}{root-raised-cosine}
\nAcronym{ROADM}{\acroSCaps{roadm}}{reconfigurable optical add-drop multiplexer}
\nAcronym{ROI}{\acroSCaps{roi}}{region of interest} 
\nAcronym{RTO}{\acroSCaps{rto}}{real-time oscilloscope} 
\nAcronym{RZDBPSK}{\acroSCaps{rz-dbpsk}}{return-to-zero differential binary phase-shift keying} 
\nAcronym{RZDQPSK}{\acroSCaps{rz-dqpsk}}{return-to-zero differential quaternary phase-shift keying}
\nAcronym{RN}{\acroSCaps{RN}}{random number}

\nAcronym{S2}{\acroSCaps{S\textsuperscript{2}}}{spatially and spectrally resolved}
\nAcronym{SA}{\acroSCaps{sa}}{simulated annealing}
\nAcronym{SamPerSym}{\acroSCaps{sps}}{samples per symbol}
\nAcronym{SBS}{\acroSCaps{sbs}}{stimulated Brillouin scattering}
\nAcronym{SCM}{\acroSCaps{scm}}{subcarrier multiplexing}
\nAcronym{SDFEC}{\acroSCaps{sd-fec}}{soft-decision forward error correction}
\nAcronym{SDM}{\acroSCaps{sdm}}{space-division multiplexing}
\nAcronym{SE}{\acroSCaps{se}}{spectral efficiency}
\nAcronym{SER}{\acroSCaps{ser}}{symbol error rate}
\nAcronym{SI}{\acroSCaps{si}}{step index}
\nAcronym{SIFMF}{\acroSCaps{si-fmf}}{\usuk{step-index few-mode fiber}{step-index few-mode fibre}}
\nAcronym{SISMF}{\acroSCaps{si-smf}}{\usuk{step-index single-mode fiber}{step-index single-mode fibre}}
\nAcronym{SLM}{\acroSCaps{slm}}{spatial light modulator}
\nAcronym{SKR}{\acroSCaps{skr}}{secret key rate}
\nAcronym{SMD}{\acroSCaps{smd}}{spatial-mode dispersion}
\nAcronym{SSC}{\acroSCaps{ssc}}{spot-size converter}
\nAcronym{SMF}{\acroSCaps{smf}}{\usuk{single-mode fiber}{single-mode fibre}}
\nAcronym{SMU}{\acroSCaps{smu}}{source measure unit}
\nAcronym{SMUX}{\acroSCaps{smux}}{spatial multiplexer}
\nAcronym{SNR}{\acroSCaps{snr}}{signal-to-noise ratio}
\nAcronym{SNU}{\acroSCaps{snu}}{shot-noise unit}
\nAcronym{SOA}{\acroSCaps{soa}}{semiconductor optical amplifier}
\nAcronym[firstplural=\usuk{states of polarization (\acroSCaps{sop})}{states of polarisation (\acroSCaps{sop})}]{SOP}{\acroSCaps{sop}}{\usuk{state of polarization}{state of polarisation}}
\nAcronym{SPM}{\acroSCaps{spm}}{self-phase modulation}
\nAcronym{SPD}{\acroSCaps{spd}}{single-photon detector}
\nAcronym{SPS}{\acroSCaps{sps}}{samples per symbol}
\nAcronym{SRS}{\acroSCaps{srs}}{stimulated Raman scattering}
\nAcronym{SSB}{\acroSCaps{ssb}}{single-sideband}
\nAcronym{SSBI}{\acroSCaps{ssbi}}{signal-signal beat interference}
\nAcronym{SSFM}{\acroSCaps{ssfm}}{split-step Fourier method}
\nAcronym{SSMF}{\acroSCaps{ssmf}}{\usuk{standard single-mode fiber}{standard single-mode fibre}}
\nAcronym{STAXT}{\acroSCaps{staxt}}{short-term average cross-talk}
\nAcronym{STL}{\acroSCaps{stl}}{swept tunable laser}
\nAcronym{SVD}{\acroSCaps{svd}}{singular value decomposition}
\nAcronym{SW}{\acroSCaps{sw}}{sequence-wise}
\nAcronym{SWI}{\acroSCaps{swi}}{swept wavelength interferometry}
\nAcronym{SP}{\acroSCaps{sp}}{setpoint}

\nAcronym{TAT}{\acroSCaps{tat}}{transatlantic}
\nAcronym{TC}{\acroSCaps{tc}}{tunable coupler}
\nAcronym{TD}{\acroSCaps{td}}{time domain}
\nAcronym{TDE}{\acroSCaps{tde}}{\usuk{time domain equalizer}{time domain equaliser}}
\nAcronym{TDFA}{\acroSCaps{tdfa}}{thulium doped-fiber amplifier}
\nAcronym{TDM}{\acroSCaps{tdm}}{time-domain multiplexing}
\nAcronym{TDMSDM}{\acroSCaps{tdm-sdm}}{time-domain multiplexed space-division multiplexing}
\nAcronym{TE}{\acroSCaps{te}}{transverse electric}
\nAcronym{TRNG}{\acroSCaps{TRNG}}{true random number generator}
\nAcronym{TEC}{\acroSCaps{tec}}{thermally-expanded-core}
\nAcronym{TOPS}{\acroSCaps{tops}}{thermo-optic phase shifter}
\nAcronym{TLS}{\acroSCaps{tls}}{tunable laser source}
\nAcronym{TIA}{\acroSCaps{tia}}{trans-impedance amplifier}
\nAcronym{TM}{\acroSCaps{tm}}{transverse magnetic}
\nAcronym{TH4D}{\acroSCaps{th-4d}}{time domain hybrid four-dimensional}
\nAcronym{TH4D2A8PSK}{\acroSCaps{th-4d-2a8psk}}{time-domain hybrid four-dimensional two-amplitude eight-phase-shift keying}
\nAcronym{TP}{\acroSCaps{tp}}{twisted pair}
\nAcronym{TTL}{\acroSCaps{ttl}}{transistor-transistor logic}

\nAcronym{UWB}{\acroSCaps{uwb}}{ultra-wideband}
\nAcronym{ULI}{\acroSCaps{uli}}{ultrafast laser inscription}

\nAcronym{VCSEL}{\acroSCaps{vcsel}}{vertical-cavity surface emitting laser}
\nAcronym{VHDL}{\acroSCaps{vhdl}}{\acroSCaps{Vhsic} Hardware Description Language}
\nAcronym{VOA}{\acroSCaps{voa}}{variable optical attenuator}

\nAcronym{WDL}{\acroSCaps{wdl}}{wavelength-dependent loss}
\nAcronym{WDM}{\acroSCaps{wdm}}{wavelength-division multiplexing}
\nAcronym{WFS}{\acroSCaps{wfs}}{wavefront sensor}
\nAcronym{WGA}{\acroSCaps{wga}}{weakly guiding approximation}
\nAcronym{WGN}{\acroSCaps{wgn}}{white Gaussian noise}
\nAcronym[longplural=wavelength selective switches]{WSS}{\acroSCaps{wss}}{wavelength selective switch}
\nAcronym{WC}{\acroSCaps{wc}}{weakly coupled}
\nAcronym{WCMCF}{\acroSCaps{wc-mcf}}{\usuk{weakly-coupled multi-core fiber}{weakly-coupled multi-core fibre}}

\nAcronym{XGM}{\acroSCaps{xgm}}{cross-gain modulation}
\nAcronym{XPM}{\acroSCaps{xpm}}{cross-phase modulation}
\nAcronym{XOR}{\acroSCaps{xor}}{exclusive or}
\langcheck{%
    \nAcronym{XPOLM}{\acroSCaps{xp}ol\acroSCaps{m}}{cross-polarization modulation}
    }{%
    \nAcronym{XPOLM}{\acroSCaps{xp}ol\acroSCaps{m}}{cross-polarisation modulation}
}%
\nAcronym{XT}{\acroSCaps{xt}}{cross-talk}

\nAcronym{3DWG}{\acroSCaps{3dwg}}{3D-waveguide}
\begin{document}
\selectlanguage{english}    


\title{ Reducing Turbulence-Induced Outages in a Deployed Terrestrial Free-Space Optical Communication Link via Interleaving \vspace{-5mm}}%


\author{
Kadir G\" um\" u\c s\textsuperscript{*}, Vincent van Vliet, Menno van den Hout, Thomas Bradley, \\ Eduward Tangdiongga, and Chigo Okonkwo
\vspace{-4mm}
}
\maketitle                  

\begin{strip}
    \begin{author_descr}

Electro-Optical Communication Group, Eindhoven University of Technology, The Netherlands

\centering
\textsuperscript{*} \textit{k.gumus@tue.nl}

\vspace{-2mm}
    \end{author_descr}
\end{strip}


\begin{strip}
    \begin{ecoc_abstract}
 We present an experimental study of data interleaving for terrestrial free-space optical communication over a 4.6~km urban testbed. Results demonstrate a two-order-of-magnitude reduction in outage probability. A dependency between measured turbulence strength, interleaver length, and achievable data rate is revealed, enabling robust system design. © 2026 The Author(s) 
\vspace{-4mm}
    \end{ecoc_abstract}
\end{strip}

\begin{figure*}[b!]
\vspace{-6mm}
    \centering
    \resizebox{0.98\linewidth}{!}{\input{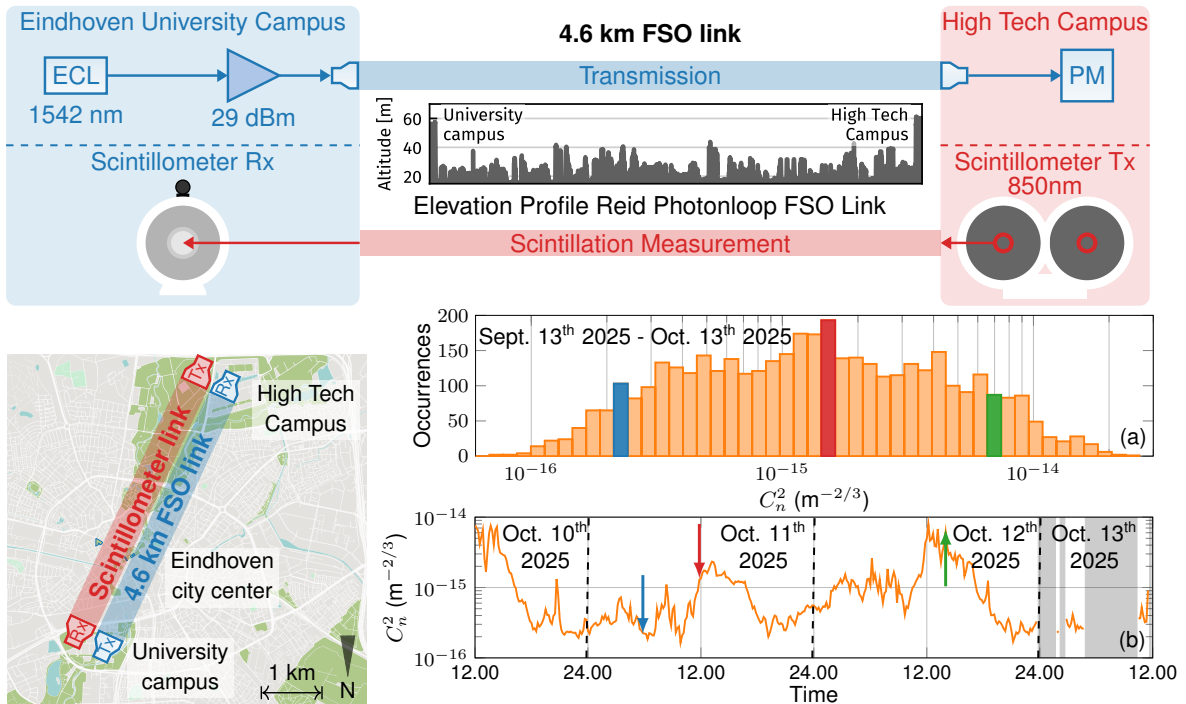}}
    \caption{Experimental setup for power and scintillation measurements over 4.6~km for a free-space optical link. The insets show (a) the distribution of measured $C_n^2$ from September 13$^{\text{th}}$ to October 13$^{\text{th}}$ 2025 and (b) the $C_n^2$ measurements from October 10${^\text{th}}$ 12.00 to October 13$^{\text{th}}$ 12.00. For insufficient received power the background is grey. The blue, red and green correspond to the investigated weak, moderate and strong turbulence cases. }
    \label{fig:setup}
\end{figure*}
\section{Introduction}%
\vspace{-2mm}%
\noindent \uFSO communication can enable high-data-rate wireless communication by leveraging technologies used in coherent optical fibre transmission. Although more commonly studied for non-terrestrial links \cite{al2020survey,liang2023free}, high-capacity terrestrial links are starting to be demonstrated experimentally in recent years\cite{vanVliet_JLT,bai2025112}. Terrestrial use cases include emergency networks, cellular backhaul, and security applications such as quantum key distribution \cite {khalighi2014survey}, which require high reliability. 

One of the main challenges in \FSO is atmospheric turbulence, which causes significant fluctuations in the system's received optical power and introduces random fluctuations in irradiance. Hence, the system's \SNR, and therefore its data rate, changes over time. For a terrestrial link, these fluctuations can be more severe than in satellite systems because the beam travels farther through the dense part of the atmosphere. This affects the reliability of terrestrial \FSO links, leading to high outage probability and preventing widespread deployment \cite{khalighi2014survey}. 

Data interleaving has been proposed as an effective mitigation technique  \cite{shi2004interleaving},  redistributing burst errors caused by turbulence across multiple codewords to improve error-correction performance. These interleavers are often used in conjunction with other link-layer solutions, such as automatic repeat request and its variants \cite{Hoang2022}, which is outside of the scope of this paper. Interleaving has been investigated extensively for satellite-based FSO systems \cite{arrieta2023proof, Ollie2026}. Although experimental works on terrestrial links do exist \cite{Greco2009,fujita2019experimental}, they do not analyse how outage probability is affected by interleavers under realistic urban atmospheric conditions.

In this work, we address this gap by analysing interleaving in a deployed terrestrial \uFSO system. Using the Reid Photonloop testbed \cite{vanVliet_JLT}, a 4.6 km urban link, we quantify how interleaver length influences outage probability under varying turbulence conditions.  Furthermore, we show a clear relationship among turbulence strength, interleaver length, and data rate, providing practical design insights for reliable terrestrial \FSO deployment.

\begin{figure*}[t!]
        \includegraphics[width = \linewidth]{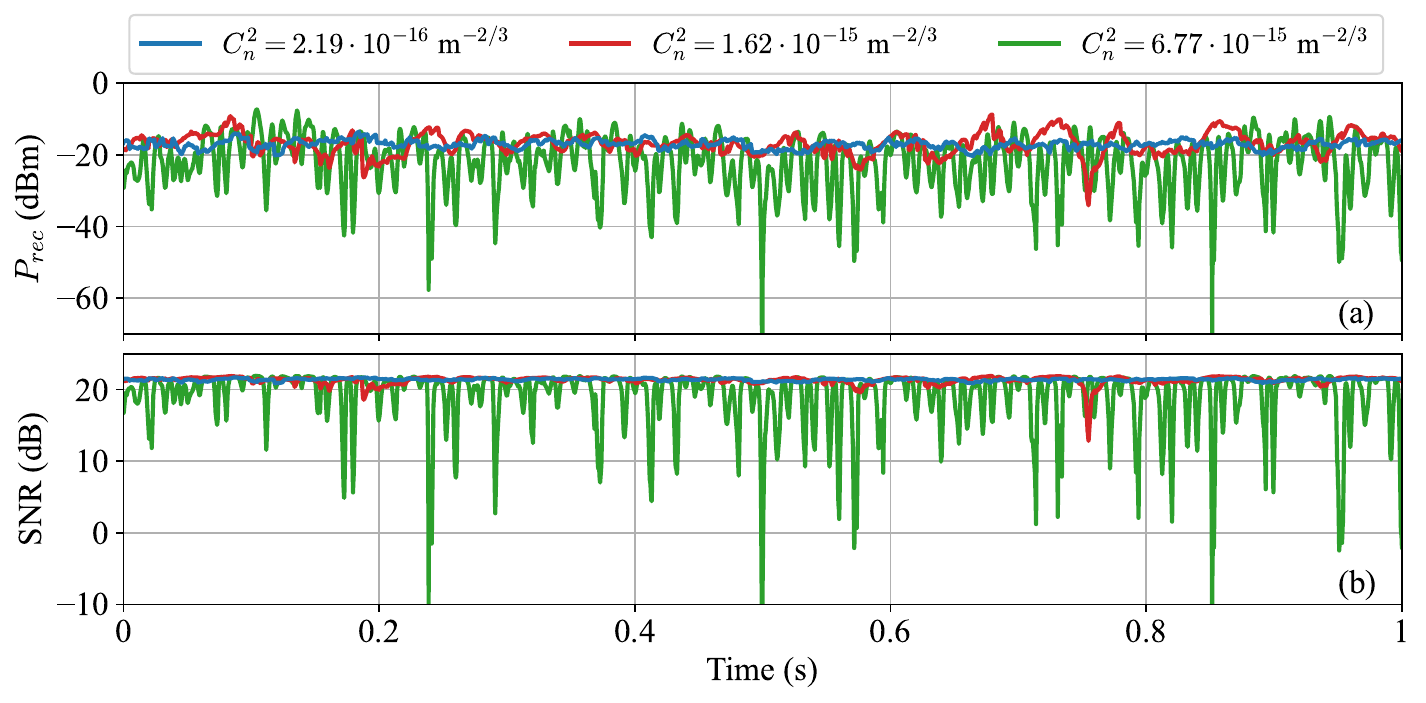}
        \caption{(a): Received optical power, $P_{rec}$ (dBm) vs time, (b) SNR (dB) vs time for 3 different turbulence cases.}
        \label{fig:Power_SNR}
        \vspace{-1mm}
\end{figure*}
\vspace{-3mm}
\section{Experimental Setup}
\vspace{-2mm}
Our terrestrial \FSO setup, as shown in Fig. \ref{fig:setup}, spans 4.6~km over the city of Eindhoven, from the Eindhoven University of Technology Campus to the High Tech Campus (HTC). To measure the received optical power $P_{rec}$ at the \FSO link receiver, we use an \ECL at 1542~nm, which is amplified by an \EDFA to 29~dBm. The light is transmitted across the city via fibre-coupled optical terminals designed by Aircision. On the HTC side, we measure the received power $P_{rec}$ using a high-speed power meter at 10\SI{}{\micro\text{s}} intervals. The power measurements ran from October 10$^{\text{th}}$ to 13$^{\text{th}}$ 2025. 

In parallel with our power measurements, we measure turbulence strength using a co-located Scintec BLS900 Neo Dual-Disk Design scintillometer. This measurement system characterises the optical turbulence in the \FSO channel using the refractive index structure parameter $C_n^2$, with an averaging time of 10 minutes\cite{vanVliet_OFC:26}. Although traversing the \FSO link in opposite directions, the scintillometer and power meter measurements do correlate. Fig. \ref{fig:Power_SNR} shows that for higher measured $C_n^2$, the $P_{rec}$ fluctuations on the \FSO receiver are more severe. This correlation was observed across our measurements.
The scintillation measurements are from September 13$^{\text{th}}$ to October 13$^{\text{th}}$ 2025.

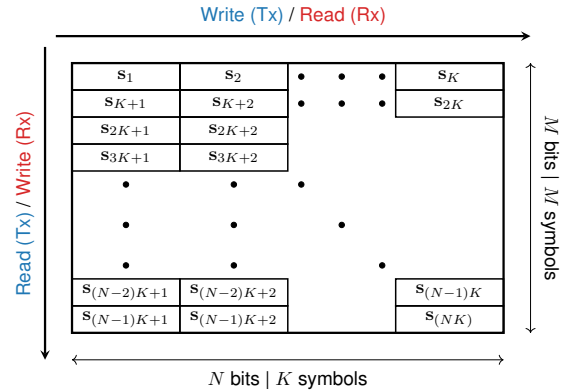
\begin{figure}[b!]
\vspace{-8mm}
    \centering
    \resizebox{0.95\linewidth}{!}{\begin{tikzpicture}
    \draw[-{stealth}, very thick] (0,5.8) -- node[above, xshift = -2mm, rotate = 90]{\textcolor{C1}{Read (Tx)} / \textcolor{C4}{Write (Rx)}} (0,0);
    \draw[-{stealth}, very thick] (0.2,6) -- node[above,yshift = 2mm]{\textcolor{C1}{Write (Tx)} / \textcolor{C4}{Read (Rx)}} (9,6);
    \draw[very thick] (0.5,5.5) rectangle (8.5,0.5);

    \foreach \x in {2,4,8}
        \draw[thick] (-1.5+\x,5.5) rectangle (0.5+\x,5);
    \foreach \x in {4,8}
        \draw[thick] (-1.5+\x,5) rectangle (0.5+\x,4.5);
    \foreach \x in {1,2,7,8,9,10}
        \draw[thick] (0.5,0.5*\x) rectangle (2.5,0.5+0.5*\x); 
    \foreach \x in {1,2,7,8,9,10}
        \draw[thick] (2.5,0.5*\x) rectangle (4.5,0.5+0.5*\x);
    \foreach \x in {1,2}
        \draw[thick] (6.5,0.5*\x) rectangle (8.5,0.5+0.5*\x);
    \foreach \x in {0,1,2}
        \filldraw (4.75 + 0.75*\x, 3.25-0.75*\x) circle (0.05cm);
    \foreach \x in {0,1,2}
        \filldraw (4.75 + 0.75*\x, 5.25) circle (0.05cm);
    \foreach \x in {0,1,2}
        \filldraw (4.75 + 0.75*\x, 4.75) circle (0.05cm);
    \foreach \x in {0,1,2}
        \filldraw (1.5,1.75+0.75*\x) circle (0.05cm);
    \foreach \x in {0,1,2}
        \filldraw (3.5,1.75+0.75*\x) circle (0.05cm);
        
    \node at (1.5,5.25) {$\mathbf{s}_1$};
    \node at (3.5,5.25) {$\mathbf{s}_2$};
    \node at (7.5,5.25) {$\mathbf{s}_{K}$};

    \node at (1.5,4.75) {$\mathbf{s}_{K+1}$};
    \node at (3.5,4.75) {$\mathbf{s}_{K+2}$};
    \node at (7.5,4.75) {$\mathbf{s}_{2K}$};

    \node at (1.5,4.25) {$\mathbf{s}_{2K+1}$};
    \node at (1.5,3.75) {$\mathbf{s}_{3K+1}$};
    \node at (1.5,1.25) {$\mathbf{s}_{(N-2)K + 1}$};
    \node at (1.5,0.75) {$\mathbf{s}_{(N-1)K + 1}$};

    \node at (3.5,4.25) {$\mathbf{s}_{2K+2}$};
    \node at (3.5,3.75) {$\mathbf{s}_{3K+2}$};
    \node at (3.5,1.25) {$\mathbf{s}_{(N-2)K + 2}$};
    \node at (3.5,0.75) {$\mathbf{s}_{(N-1)K + 2}$};

    \node at (7.5,1.25) {$\mathbf{s}_{(N-1)K}$};
    \node at (7.5,0.75) {$\mathbf{s}_{(NK)}$};
    
    \draw[<->] (9,0.5) -- node[midway,above,xshift = 2mm,rotate = -90]{$M$ bits | $M$ symbols}(9,5.5);

    \draw[<->] (0.5,0) -- node[midway,below,yshift = -2mm]{$N$ bits | $K$ symbols}(8.5,0);
\end{tikzpicture}}
    \caption{Overview of a symbol-wise block (de)interleaver.}
    \label{fig:interleaver}
\end{figure}

In the insets in Fig. \ref{fig:setup} (a), the distribution of $C_n^2$ for the period of one month and (b) $C_n^2$ measured in the same period as the optical power measurements are shown. For 4.2\% of the scintillation measurements, the received optical power was below the threshold for the scintillometer receiver to measure $C_n^2$. These measurements are not taken into account in inset (a) and are shown with a grey background in inset (b). The distribution of $C_n^2$ is centered around $1.5\cdot10^{-15}$~m$^{-2/3}$, with two more peaks around $5\cdot10^{-16}$~m$^{-2/3}$ and $4\cdot10^{-15}$~m$^{-2/3}$, caused by a higher level of atmospheric turbulence during the day, and lower at night, as shown by the diurnal pattern indicated in inset (b).

\begin{figure*}[t!]
        \includegraphics[width = \linewidth]{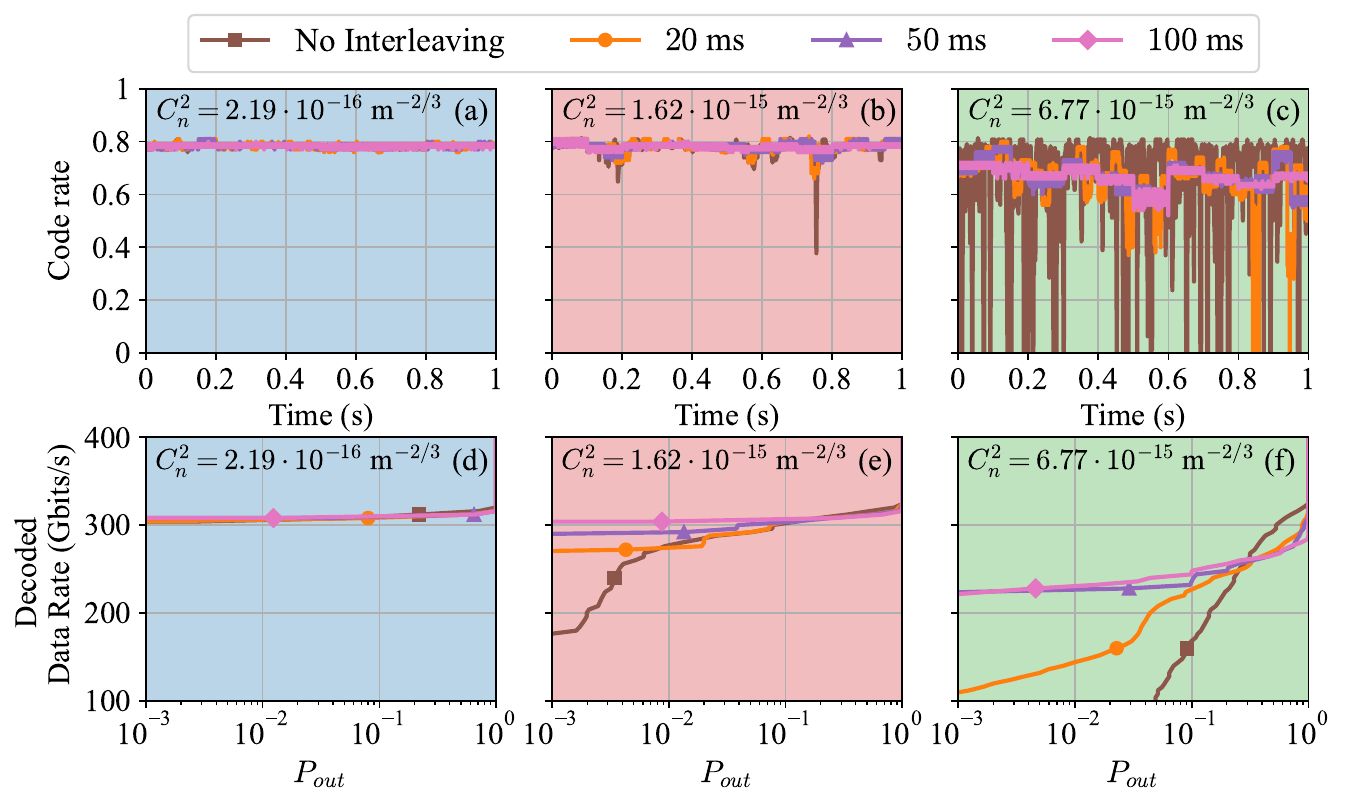}
        \caption{(a, b, c) Code rate vs time, (d, e, f) Throughput vs $P_{out}$ for four different interleaver lengths. (a,d): weak turbulence, (b,e): moderate turbulence, (c,f): strong turbulence.}
        \label{fig:Rate_Pout}
        \vspace{-1mm}
\end{figure*}

To investigate the effect of bit interleaving on the outage probability $P_{out}$, the probability that the decoded code rate is lower than the operational code rate, we take a one second sample from our power meter results corresponding to weak (blue, $C_n^2 = 2.19\cdot10^{-16}$~m$^{-2/3}$), moderate (red, $C_n^2 = 1.62\cdot10^{-15}$~m$^{-2/3}$, and strong (green, $C_n^2 = 6.77\cdot10^{-15}$~m$^{-2/3}$) observed turbulence. These samples were chosen based on the distribution of $C_n^2$ shown in Fig. \ref{fig:setup}(a), picking samples in the middle, and toward the tail ends of the distribution. The received power $P_{rec}$ for these traces is shown in Fig. \ref{fig:Power_SNR} (a). We assume the following for our system: We use an \EDFA with a noise figure of 4.4~dB to amplify the received signal to 0~dBm. To calculate the ASE noise generated by the \EDFA, we use eq. (54) from \cite{essiambre2010capacity}. The symbol rate $B$ of our system is 50~GBaud, using 256-QAM. The $\text{SNR}_{B2B}$ we measured of our back-to-back system is 22~dB. The total SNR, as shown in Fig. \ref{fig:Power_SNR} (b), is $\text{SNR} = \frac{1}{1/\text{SNR}_{B2B} + 1/\text{SNR}_{EDFA}}$, where $\text{SNR}_{EDFA}$ is the SNR of a shot-noise-limited signal amplified by the receiver \EDFA. Both $P_{rec}$ and the SNR fluctuate more for increasing turbulence.
\vspace{-3mm}
\section{Decoding with Interleaver}
\vspace{-2mm}
For the error correction, we use punctured \LDPC codes from the DVB-S2 standard \cite{morello2006dvb}, which have an error correction blocklength $N$ of 64800 bits. The minimum rate for the DVB-S2 standard is $\frac{1}{4}$; hence, whenever a lower rate is required, decoding fails and there is no throughput.  The block (de)interleaver used in the simulation is shown in Fig. \ref{fig:interleaver}. The size of the interleaver depends on $N$, the length of the interleaver in seconds $t$, and the symbol rate $B$. The interleaver has $K$ columns, where $K = \frac{N}{m}$ and $m$ is the number of bits per symbol $\mathbf{s}_i$, and $M = \frac{tB}{N}$ rows. On the transmitter side, we write our symbol stream into the interleaver row by row. After filling the interleaver matrix, we read out the symbols column by column and transmit them over the channel in this order. At the receiver side, we do the exact opposite: write to the deinterleaver column by column and read out row by row, returning to the original symbol order. Any burst errors caused by fast fading are now spread over multiple codewords. However, the interleaver introduces a latency equal to $2t$, as the interleaver and deinterleaver need to be completely filled before transmission and decoding can start, respectively. For the error correction simulation, we simulate per 10 \SI{}{\micro\text{s}} time instance down to an outage probability $P_{out}$ of $10^{-3}$.  
In Fig. \ref{fig:Rate_Pout}, we show the code rate vs time and the throughput vs $P_{out}$ for all 3 investigated turbulence scenarios and different interleaver lengths. In the low-turbulence case, throughput is consistent over time, and interleaving does not significantly affect $P_{out}$. In the moderate case, interleaving allows for up to a 70\% higher decoded data rate compared to no interleaving when $P_{out} < 10^{-2}$ due to the deep fade at 0.75 s. The 20 ms interleaver significantly reduces the outage probability, with longer interleaver lengths yielding incremental gains in decoded data rate. For the high-turbulence case, an interleaver of at least 50 ms is required to get $P_{out} < 10^{-2}$ at decoded data rates above 200Gb/s. These results show that as $C_n^2$ increases, the channel becomes less stable, and longer interleavers are required for high data rates at low $P_{out}$, at the cost of higher latency.
\vspace{-3mm}
\section{Conclusion}
\vspace{-2mm}
This work experimentally characterised the distribution of $C_n^2$ over a 4.6km terrestrial FSO link and quantified its impact on outage probability. We show that interleaving is a decisive factor in mitigating turbulence-induced outages, with optimal interleaver length strongly dependent on turbulence strength. In particular, increasing $C_n^2$ necessitates longer interleavers to sustain high data rates at low outage probabilities, in our case $P_{out} < 10^{-2}$ for data rates $\geq 200$~Gb/s, at the cost of higher latency. These results establish a direct link between atmospheric conditions and interleaver design, providing a foundation for adaptive interleaving strategies in practical terrestrial FSO systems.

\clearpage
\section{Acknowledgements}
Supported by the Dutch Research Council (NWO) TTW-Perspectief Optical Wireless Superhighways: Free photons (at home and in space): FREE P19-13, the Dutch Ministry of Economic Affairs and Climate Policy (EZK) via the PhotonDelta National Growth Fund Programme on Photonics, and European Innovation Council Transition project CombTools (G.A. 101136978). We thank Aircision B.V., particularly Nourdin Kaai, Luis Pellicer Collado, and Roland Blok, for their support of the Reid Photonloop \FSO testbed, and Keysight Technologies for providing the high-speed power meter.


\printbibliography
\end{document}
